\documentclass[%
 reprint,
superscriptaddress,
 amsmath,amssymb,
 aps,
prl,
noeprint
]{revtex4-2}

\usepackage{bbold}
\usepackage{graphicx}
\usepackage{dcolumn}
\usepackage{bm}
\usepackage[colorlinks=true,linkcolor=blue,citecolor=blue,urlcolor=blue]{hyperref}

\newcommand{\Methods}{Supporting Information}
\newcommand{\Supp}{Supporting Information}

\newcommand{\Qout}{Q^{\mathrm{(out)}}}
\newcommand{\Qin}{Q^{\mathrm{(in)}}}

\begin{document}

\preprint{APS/123-QED}

\title{Active fluid networks excite visco-elastic modes for efficient transport}

\author{Adam B. Dionne}
\email{adamdionne@g.harvard.edu }
\affiliation{%
Physics Department, Williams College, 33 Lab Campus Drive, Williamstown, MA 01267, U.S.A.
}%
\affiliation{%
School of Engineering and Applied Sciences, Harvard University, Cambridge, MA 02138, U.S.A.
}%
\author{Katharine E. Jensen}
\author{Henrik Ronellenfitsch}%
 \email{henrik.ronellenfitsch@gmail.com}
\affiliation{%
Physics Department, Williams College, 33 Lab Campus Drive, Williamstown, MA 01267, U.S.A.
}%

\date{\today}

\begin{abstract}
Active fluid transport is a hallmark of many biological transport networks. While 
animal circulatory systems generally rely on a single heart to drive flows, 
other organisms employ decentralized local pumps to distribute fluids and nutrients. 
Here, we study the decentralized pumping mechanism
in the slime mold \emph{Physarum polycephalum} which is locally triggered by 
active release, uptake, and transport of a chemical solute within the 
organism's vascular network to drive global oscillations.
Based on a conceptual network model combining active elasticity and 
fluid transport we identify a set of contractile modes specific to each network and show that modes corresponding
to large-scale oscillations are preferentially and robustly excited both 
in model simulations and 
in experimental data obtained from living \emph{Physarum} plasmodia. These dominant
modes are computed explicitly and shown to drive large-scale flows within the
organism. 
Furthermore, \emph{Physarum} must transport nutrients over long distances.
As each mode corresponds to pure shuttle flow, long-range, directed transport
must rely on a non-linear coupling beyond harmonic dynamics. Using simulations,
we demonstrate that the network's transport capability is optimized when two 
dominant modes are excited at a phase shift of $\pi/2$, resulting in contractile
excitations similar to those observed in real \emph{Physarum}.
Our results provide a conceptual framework for understanding active 
decentralized transport in \emph{Physarum} and other contractile 
biological networks, such as brain vasculature, as well as decentralized transportation
networks more generally.
\end{abstract}

\maketitle

\section{Introduction}
Active transport networks are ubiquitous in nature and engineering.
Gas and sewage networks are engineered for efficient delivery of fluids in large-scale
production~\cite{Osiadacz1996} and infrastructure~\cite{Cunha1999} contexts, and may be 
centrally driven.
The animal vascular system is actively driven by one central heart in the case of mammals~\cite{Nguyen2021} or several
in the case of insects~\cite{Hillyer2020} and 
is used for long-range  transport of oxygen and nutrients. 
Plants employ decentral actively maintained concentration 
gradients for sugar transport in the phloem network~\cite{Katifori2018,Jensen2012,DeSchepper2013}, and contractile muscles 
surrounding the brain vasculature are recruited by surrounding neurons for localized 
oxygen delivery~\cite{Ross2020}. Thus, while engineered networks are often centrally driven,
decentralized local driving appears to be an effective strategy in biological transport. Furthermore, both biological and
human-made networks can often be described in the language
of synchronizing coupled oscillators~\cite{Kapral1995,Strogatz2012},
and many examples are optimized for a particular 
functionality~\cite{Rocks2019,Ronellenfitsch2016,Gavrilchenko2021} and can
adapt to their environment~\cite{Hu2013,Fricker2009,Ronellenfitsch2019} by
harnessing the combined effects of fluids and elasticity~\cite{Luo2022}.

Here, we focus on the well-studied biological model organism 
\emph{Physarum polycephalum}~\cite{Alim2013,Oettmeier2017,Oettmeier2020}, a slime mold which in its plasmodium phase
consists of a complex network of contractile tubes filled with cytoplasm~\cite{Goldstein2015}. 
The organism is able to actively contract and expand its tubular network
using a chemical trigger, producing
large scale oscillatory waves and excitations that drive flows and transport
nutrients across its body~\cite{Alim2017,Teplov2010,Wohlfarth-Bottermann1979,Smith1992,Teplov2017}. 
It has been shown that \emph{Physarum} is able to self-organize and react
to external stimuli~\cite{Ueda1976,Alim2018}, solve complex computational 
problems~\cite{Tero2008,Tero2010,Nakagaki2000,Adamatzky2009},
optimize its transportation 
network~\cite{Tero2010,Alim2013a,Marbach2016,Bauerle2020,Marbach2023} 
and store memories based on available food~\cite{Kramar2021}.
An important emerging class of models to describe 
biological dynamics, empirical mode 
decompositions~\cite{Fujii2020,Cohen2022,Romeo2021,HTu2014}, have been applied to \emph{Physarum} data, revealing a continuous spectrum of empirical modes~\cite{Fleig2022}.
While significant effort has been made to study the local properties of
\emph{Physarum}'s chemically driven visco-elastic active oscillations~\cite{Alim2017,Julien2018} as well as the
phenomenology of the network oscillations
~\cite{Alim2017,Alim2013a},
a conceptual understanding of the entire network dynamics and its relationship
to functionality is still missing.

\begin{figure*}[t!]
    \centering
     \includegraphics{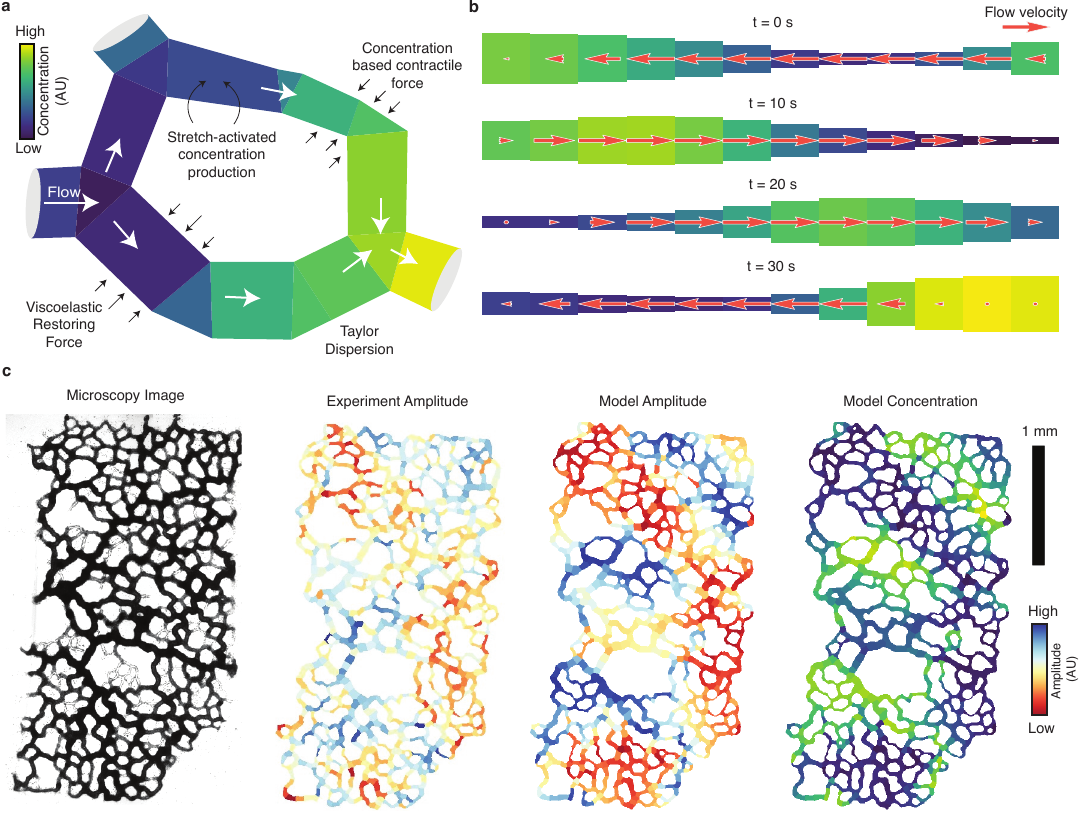}
    \caption{Our visco-elastic flow network model. 
    \textbf{a}, A cartoon of the model which depicts pressure driven flow (white arrows) travelling through discretized cylindrical vessels. A contraction triggering chemical travels through the vessels, and is released when vessels are stretched past their equilibrium. \textbf{b}, The model implemented on a $n=12$ path graph showcases a peristaltic wave of contractions. Still frames taken from Video 1, where the red arrows depict flow velocity. \textbf{c}, The model implemented on a real \textit{Physarum polycephalum} network, corresponding to the microscopy image. The experimentally observed large-scale organization of contraction amplitudes are reproduced by the active network model, which is based on contraction-triggering solutes. Microscopy and model network still frames taken from Videos 2 and 3 respectively.}
    \label{fig:fig1}
\end{figure*}

Here, we propose a network model of visco-elastic active oscillations
and fluid flow in \emph{Physarum} based on physical principles and show using simulations that it can
reproduce large-scale coordinated contraction waves similar to those observed in
real \emph{Physarum} networks. Based on analysis of our model, we then show that
it is possible to compute a set of linear contractile modes based on 
network topology alone. We demonstrate that only a few of the largest-scale 
modes are
sufficient to explain the network dynamics in both our simulations and in
experimental data obtained from analysis of brightfield microscopy 
time series of \emph{Physarum} contractions.
We further show that the model dynamics corresponds to an attempt to 
maximize order and synchrony among the individual pumping
vessel elements, subject to physical constraints.
Finally, we employ the conceptual knowledge gained to study the \emph{Physarum}
network's functionality: long-range transport of nutrients. Using a simple
model based on advection of particles on top of the network fluid flows, 
we demonstrate in real \emph{Physarum} network topologies that 
long-range transport is maximized when the two largest modes are excited at
a phase shift of approximately $\pi/2$. 
This phase shift had been measured but not fully explained in previous experiments~\cite{Bauerle2020}. 

\section{Results and Discussion}

\begin{figure*}[t!]
    \centering
     \includegraphics{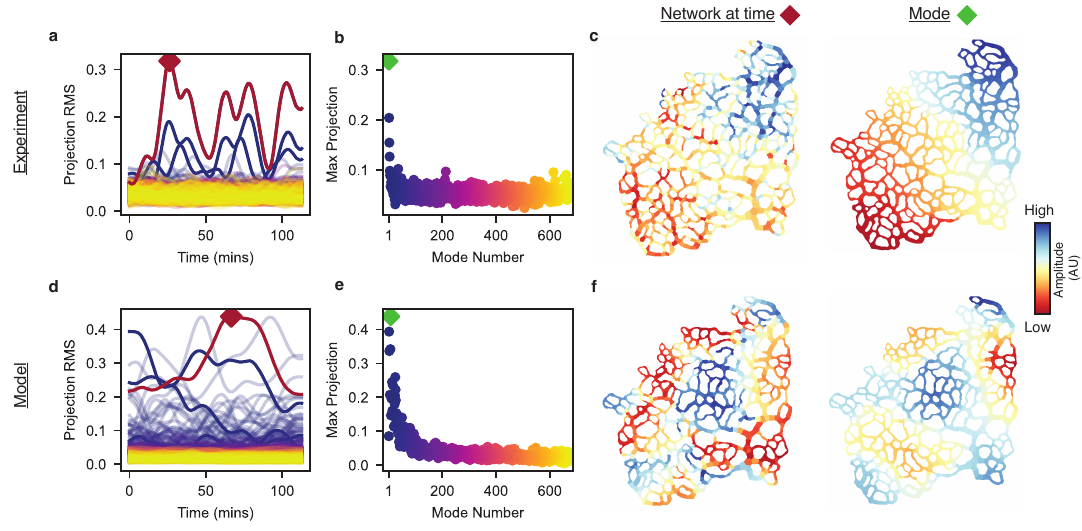}
    \caption{Both the observed and modeled temporal dynamics of \emph{Physarum} can be decomposed by the network's visco-elastic decay modes. \textbf{a}, The observed \textit{Physarum} dynamics is projected onto each mode, and the root-mean-squared signal of that projection is plotted over time. Each mode's corresponding curve is colored based on its mode number. The mode with the maximum projection is plotted in red, and the time where this maximum is achieved is denoted by the diamond. \textbf{b}, A plot of each mode's maximum projection, where the lowest modes are colored in yellow and the highest in purple. The mode with the largest projection is denoted by the red diamond. \textbf{c}, The mode with the largest projection is displayed on the right, and the observed \textit{Physarum} behavior when the maximum is achieved is on the left. \textbf{d-f}, The same analysis as \textbf{a-c} but for modeled \textit{Physarum} dynamics.}
    \label{fig:modes}
\end{figure*}

In its plasmodium stage, \emph{Physarum} consists of a densely connected,
hierarchically organized network of cytoplasm-filled tubes which are able to 
rhythmically contract, leading to shuttle flow of the cytoplasm~(Fig.~\ref{fig:fig1} (\textbf{a}, 
\textbf{b})).
The mechanism by which the contractions synchronized across the
entire organism is through fluid advection of a triggering chemical,
generally believed to be Calcium ions~\cite{Smith1992}. 
The walls of \emph{Physarum}'s tubes are embedded with an active
actomyosin cortex that reacts to the presence of Calcium ions by contracting, 
and with Calcium channels that release Ca$^{2+}$ ions upon mechanical extension~(Ref.~\cite{Veksler2009}, Fig.~\ref{fig:fig1}).
Together, these two processes lead to self-sustained sinusoidal 
oscillations of the tube walls and with that to fluid flow inside
the network. The flow is able to transport the triggering chemical,
leading to long-range organization across the organism (Fig.~\ref{fig:fig1}
(\textbf{c})). The active elasto-chemical oscillations have been studied in 
models~\cite{Veksler2009,Smith1992}. We now consider a model that extends
to an entire complex network topology, based on the one-dimensional
model from Ref.~\cite{Julien2018}.
It turns out to be useful to proceed in two steps. We first construct
a passive model coupling fluid transport and visco-elastic forces
within the vessel walls. This model already yields a useful modal
description of the network dynamics.
In a second step, we include solute-triggered contractions and solute
dynamics, capturing the active, self-sustained oscillations observed
in \emph{Physarum} plasmodia.

We model the biological vascular network as a graph, with each edge
$(ij)$ directed from node $j$ to node $i$ representing
one entire tube, or a piece of a tube of volume $V_{ij}$. Because the
tube volume is changing at a rate of $\dot{V}_{ij}$, 
the volume flow rates into and out of each vessel are related by
\begin{align}
    \Qout_{ij} = \Qin_{ij} - \dot{V}_{ij}. \label{eq:inout}
\end{align}
Here, the volume flow rates $Q_{ij}$ are positive if flow is directed 
from node $j$ to node $i$, and negative otherwise. At each junction of the network,
the total flow must be conserved,
\begin{align}
    \sum_j \Qin_{ij} = 0, \label{eq:junct-conservation}
\end{align}
and over the time scales we consider, the total volume of the network
is approximately constant,
\begin{align}
    \sum_{(ij)}\dot{V}_{ij} = 0. \label{eq:netw-conservation}
\end{align}
Following Ref.~\cite{Julien2018}, we choose a visco-elastic model
for the mechanical properties of the network. Each vessel 
acts as an overdamped spring following
\begin{align}
    \tau_{ij}\dot{V}_{ij} = E\varepsilon_{ij} + \sigma_{ij},
    \label{eq:vessel-elastic}
\end{align}
where $\tau_{ij}$ is the damping constant, $E$ is the Young's modulus, the strain $\varepsilon_{ij} = (R_{ij} - R^{(0)}_{ij})/R^{(0)}_{ij}$
with the vessel radius $R_{ij}$ and equilibrium radius $R^{(0)}_{ij}$, and $\sigma_{ij}$ is an additional external stress.
The constraints Eq. ~\eqref{eq:inout}, \eqref{eq:junct-conservation}, and \eqref{eq:netw-conservation} can be combined with the visco-elastic
force law Eq.~\eqref{eq:vessel-elastic} using a generalization of
Thompson's principle for electrical networks (\Supp, Ref.~\cite{Bollobas1998}).
In vector notation, the resulting equations of motion read
\begin{align}
    M\, \dot{\mathbf{V}} = -E \boldsymbol{\varepsilon}
    +\boldsymbol{\sigma} - \mu\mathbf{1}, \label{eq:passive}
\end{align}
where $\mathbf{V}$ is the vector of vessel volumes, 
the damping matrix $M=\tau - \frac{1}{4} B^\top L^\dagger B$
with $B$ the graph's unweighted incidence matrix and $L$ the graph's weighted
Laplacian matrix (\Methods), and $\mu$ is a Lagrange multiplier ensuring 
total volume
conservation and playing the role of an external hydrostatic pressure. 
The Laplacian encodes the radii of all vessels, assumed to be constant
in a linear approximation, by way of modeling the vessels as cylindrical
tubes subject to Poiseuille flow (\Methods).
The dagger represents the Moore-Penrose pseudoinverse, and we
introduced the vector of all ones, $\mathbf{1} = (1, \dots, 1)^\top$.
The damping matrix $M$ encodes the visco-elasticity of the vessel
walls in addition to viscous damping in the internal fluid. Here, the the pressures were eliminated.
This equation of motion
for the passive network Eq.~\eqref{eq:passive} admits solutions
that can be decomposed into modes. The volume rate of change of cylindrical vessels with length $L$ can be related to their radii $R$ by
$\dot{V} = 2\pi L R \dot{R}$. Then 
assuming small excitations around the equilibrium
radii of the form $\mathbf{R}(t) = \bar{\mathbf{{R}}} + \mathbf{u}\, e^{-\lambda t}$
leads the to generalized eigenproblem
\begin{align}
    \label{eq:eigenproblem}
    A \mathbf{u} = -\lambda M \mathbf{u},
\end{align}
where the matrix $A$ encodes vessel geometry
(\Supp).

Equation \eqref{eq:eigenproblem} provides a powerful framework
for thinking about the dynamics of \emph{Physarum} and similar systems.
Due to its mathematical properties (\Supp), Eq.~\eqref{eq:eigenproblem} is
subject to a generalization of Courant's nodal domain 
theorem~\cite{Courant1923,Urschel2018} which states that, when
arranged in ascending order of the eigenvalue, the
eigenvectors corresponding to the $n$'th eigenvalue partition the
network into no more than $n$ connected domains according to
the sign of the entries of $\mathbf{u}$. Thus, we expect
that large scale excitations of the biological network can be expressed
economically using a few of the lowest modes encoded by
Eq.~\eqref{eq:eigenproblem}. 

We now proceed to apply the modal decomposition to measured \emph{Physarum} dynamics.
Plasmodia of \emph{Physarum} were cultured on agar plates based on the
protocol in Ref.~\cite{Julien2018}, and microscopically imaged over the span
of approximately two hours to capture many periods of shuttle flow 
oscillation while
avoiding behaviors such as growth that occur over longer time spans
(\Methods). Images were then discretized using custom software, and
a graph representation of the vascular network including
the radii of all vessels over time (Fig.~\ref{fig:fig1}) was constructed
from which the modes Eq.~\eqref{eq:eigenproblem} were calculated
using the equilibrium radii (\Methods).
The time series of the $i$'th mode amplitude was computed by projecting the
vector of all vessel radii $\mathbf{R}(t)$ onto the mode,
$a_i(t) = \mathbf{R}(t) \cdot \mathbf{u}_i$. With this, the dynamics
of the network is expanded as $\mathbf{R}(t) = \sum_i a_i(t) \mathbf{u}_i$.
It can be seen that over the duration of our observations of
several specimens, a few modes dominate the dynamics,
with the bulk making only small contributions (Fig.~\ref{fig:modes}
\textbf{a-b}, \textbf{d-e}). Visualizing these dominant modes, we see that they are
large-scale excitations that span the entire network and that
they correspond to the largest eigenvalues of Eq.~\eqref{eq:eigenproblem}
(Fig.~\ref{fig:modes} \textbf{c,f}).

While much insight can be gleaned from the passive features of the organism, \emph{Physarum}'s network actively self-organizes its dynamics.
The basic model from Eq.~\eqref{eq:passive} can be augmented with
active contraction triggering by chemical solutes as well as solute
transport. We employ a discrete version of the advection-diffusion
equation to model transport, adsorption, and decay of a solute with total
amount $\mathcal{C}_{ij}$ in edge $(ij)$ of the network (\Methods). 
We again follow Ref.~\cite{Julien2018}
for the active, solute-dependent stress $\boldsymbol{\sigma}(\mathcal{C})$ (\Methods).
Large solute concentrations within a vessel trigger active contractions,
which in turn wash solute away. At the same time, relaxation of the 
vessel triggers enhanced solute production. Locally, this leads
to a self-sustained oscillation of the vessel but on large scales, transport
of the solute by advection and diffusion organizes and synchronizes
the network~\cite{Alim2017}. Similar non-linear
wave propagation phenomena have been described recently in generic models
of active transport networks~\cite{Fancher2022,Gounaris2021,Ruiz-Garcia2021},
and are known from the theory of excitable media~\cite{Meron1992}.
Most physical parameters of the model have been determined before,
but for the two parameters describing the nonlinear excitations,
no experiments exist. We therefore performed a detailed numerical parameter
study to identify the regions in parameter space where the non-linear
equations yield stable oscillatory dynamics (\Supp).
There exist parameters where the stable dynamics reproduces amplitudes 
and time scales of our \emph{Physarum} specimens (\Supp, Fig.~\ref{fig:fig1}),
suggesting that the model can match the real organism in relevant
metrics.
In this biological regime, the model predicts both contractile waves
and chemical waves of solute concentration that propagate across
the organism (Fig.~\ref{fig:fig1}).

\begin{figure*}[t!]
    \centering
    \includegraphics{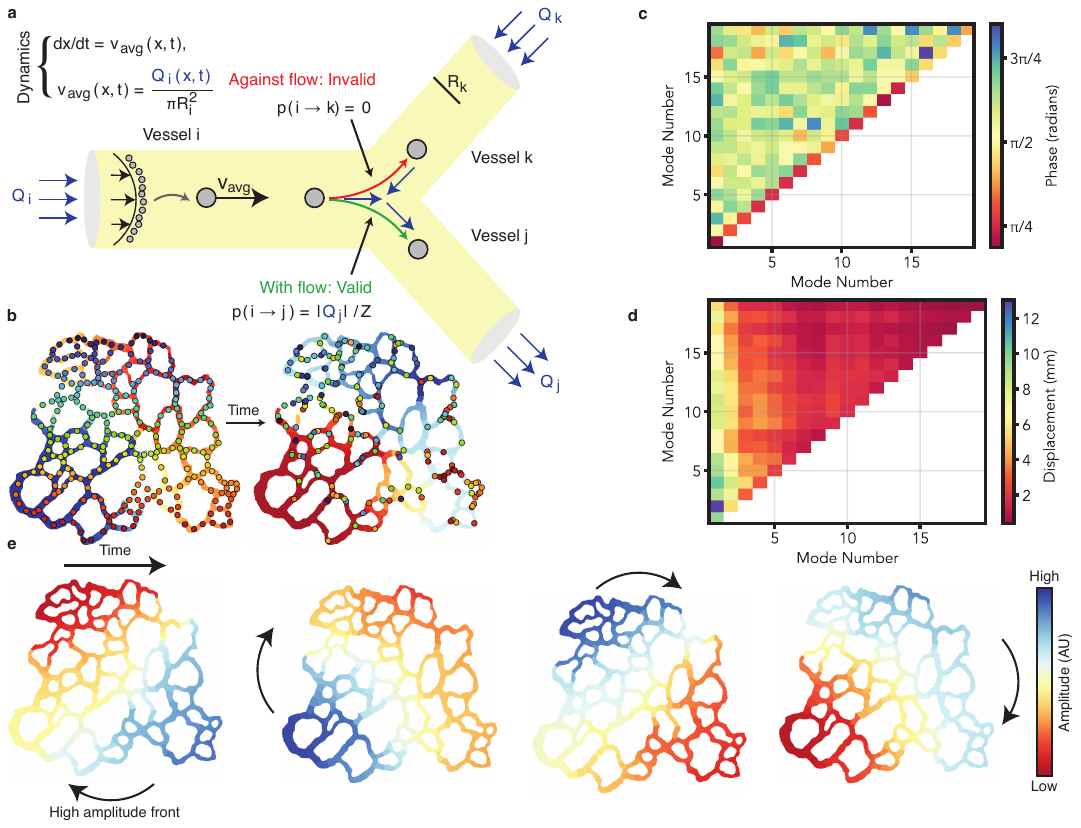}
    \caption{Nutrient transport through a \textit{Physarum} network is optimized by exciting the two dominant visco-elastic decay modes at a $\phi = \pi/2$ phase shift. \textbf{a}, An illustration of our nutrient transport model. We simulate advection dominated particles travelling at their vessel's average velocity $v_{\text{avg}}$. Junction transitions are determined stochastically, such that flow direction is preserved and transition rates are proportional to the volumetric inflow $Q$ of the target vessel: $p(i \rightarrow j) = |Q_j|/Z$, where $Z = \sum_{\text{valid }j}|Q_j|$ is the normalization. 
    \textbf{b}, An illustration of nutrient particles advected on a network, taken from Video 4. After a long time, each particle (identified by color) is transported away from its initial location. 
    \textbf{c}, We simulate transport for every pair of twenty modes with the highest excitation, and plot the optimal phase between the two modes. When overlaying two different modes, we find that $\phi = \pi/2$ optimizes transport. When overlaying the same mode, transport is optimized by no phase shift. \textbf{d}, We plot the average displacement at the end of each simulation for every mode pair phase shifted at the optimal shift. We see that overlaying large scale modes yields the best transport. \textbf{e}, Overlaying the two largest scale modes at a $\phi=\pi/2$ phase shift results in a circularly propagating contraction wave.}
    \label{fig:transport}
\end{figure*}

Armed with the non-linear model from above and an understanding
of the importance of large scale visco-elastic modes for the
dynamics of \emph{Physarum}, we now proceed to study the functionality
of the network.
Presumably, an important function of the transport network is
to take up nutrients from food sources and distribute them
throughout the organism. To reveal how dynamical modes
and transport may play together to enhance nutrient dispersion,
we simulate a simple model of transport on one of
our realistic networks. We assume that transport is
dominated by the advection of particles that move
at the velocity of the surrounding fluid. Given that 
the distribution of flow velocities in our Poiseuille 
approximation is parabolic as a function of distance from
the vessel walls (\Supp), for simplicity we consider only
particles that move with the mean velocity.
While the position $x(t)$ of a particle along each vessel 
can be found by simply integrating $\dot{x} = v(t)$, 
where $v(t)$ is the local flow velocity, at each junction
a decision must be made as to which vessel the particle will
continue in. Since a full fluid-dynamical simulation of the
junction flows is beyond the scope of our analysis, we resort
to the simple rule that the probability of choosing a
particular next vessel is proportional to the flow rate in it
(Fig.~\ref{fig:transport} \textbf{a}).
We then place three particles on each edge of the network and
simulate a combination of two elastic modes $i$ and $j$
such that the flow rates follow $\mathbf{Q}(t) = \mathbf{Q}_i(t)
+ \mathbf{Q}_j(t - \phi\, T/(2\pi))$, where $T$ is the
period and $\phi$ is a phase shift. We choose to consider two modes
because there have been indications that such combinations may be 
relevant in \emph{Physarum} organisms~\cite{Bauerle2020}.

For each combination of modes and phase shifts we then computed
the mean square displacement $\sqrt{\langle x^2\rangle}$ averaged
over both a simulation time of one hour and over all particles
(\Methods).
The mean square displacement quantifies the ability of the
network to transport nutrients quickly 
(Fig.~\ref{fig:transport}~b). For each combination of modes
we typically find that the displacement is maximized for 
a range of phase differences centered around $\pm \pi/2$,
independent of the particular pair of modes we considered
(Fig.~\ref{fig:transport}~\textbf{b}). This phase lag between the dominant 
harmonics has been observed in real 
\emph{Physarum} \cite{Bauerle2020}. The optimal phase lag 
generally corresponds to contractile waves that
move circularly around the network (Fig.~\ref{fig:transport}~\textbf{e}).
Finally, we find that the optimal combination of modes
for transport consists of the two largest modes 
(Fig.~\ref{fig:transport}~\textbf{d}). Thus, the observed organism-spanning
contractions in real networks are consistent with an adaptive
optimization for efficient transport.





\section{Conclusions}

Here we demonstrated that the dynamics of living organisms
that transport fluid and nutrients by rhythmic contraction
of tubes can be understood in terms of dynamical modes that are
based on a physical model of the organism's transport network.
Based on experimental observations of \emph{Physarum} networks over
several hours, we showed that indeed, the slime mold network
excites few large scale visco-elastic modes during its activity.
The visco-elastic modes can be computed from network topology, geometry, and
elasticity alone, without recourse to a detailed model of the
active motion. Thus, they may be applicable to more organisms than
\emph{Physarum} alone, as long as the passive physics are known.
We then proceeded to further elucidate \emph{Physarum} dynamics by
constructing a network model of its active dynamics. 
We were able to reproduce not only crucial qualitative properties of the real
dynamics like large-scale excitations, but also  
matched amplitudes and periods of radial oscillations, as well as
an order parameter of the network. 
With this complete model, we proceeded to study the functionality
of fluid-carrying networks. By considering the motion of particles
that are advected by the flows inside the network, such as particles of food
taken up by the slime mold, we demonstrated that the most effective
transport is achieved by combining the two largest-scale dynamical
modes at at relative phase of $\pi/2$. This had been seen in real
\emph{Physarum} before, suggesting that the slime mold may indeed optimize
its rhythmic contractions for nutrient transport.

While we showed 
that the dynamics of \emph{Physarum} can be
understood in terms of an interplay between a small number of 
modes determined by passive physics and network
topology, active contractions, and optimization for transport, 
we anticipate that 
our results will 
have wider-ranging consequences. 
Specifically, our passive model applies to all elastic, fluid filled
networks such as the vasculature of mammals or insects 
wings~\cite{Pass2018}. The
visco-elastic modes we computed may play a role in their passive dynamics
as well. Furthermore, there exist active fluid-pumping networks such as
the brain vasculature, where nerve cells actively control vascular 
contractions to direct oxygen transport~\cite{Huo2014}. Here, our model may have even 
more direct applicability to the study of brain hemodynamics, bridging
the gap between disparate biological systems.

\begin{acknowledgments}
A.B.D. acknowledges support from the Williams College Science Center.
K.E.J. and H.R. each acknowledge support from Williams College startup grants.

\end{acknowledgments}



\bibliography{mainbib}

\end{document}



\title{Supporting Information for \\``Active fluid networks excite visco-elastic modes for efficient transport"}

\author{Adam B. Dionne}
\affiliation{%
Physics Department, Williams College, 33 Lab Campus Drive, Williamstown, MA 01267, U.S.A.
}%
\affiliation{%
School of Engineering and Applied Sciences, Harvard University, Cambridge, MA 02138, U.S.A.
}%
\author{Katharine E. Jensen}
\author{Henrik Ronellenfitsch}%
 \email{henrik.ronellenfitsch@gmail.com}
\affiliation{%
Physics Department, Williams College, 33 Lab Campus Drive, Williamstown, MA 01267, U.S.A.
}%


\maketitle

\beginsupplement

\tableofcontents

\section{Full Network Model}
\subsection{Fluid Transport Network}
We begin by formalizing a network of cylindrical tubes that pump fluid through rhythmic contractions and expansions. To do so we leverage Thomson's principle \cite{Bollobas}, which asserts that physical flows minimize dissipated power under the constraint that flows are conserved. Let's assume contractions are slow such that the flow remains steady, laminar, and fully developed. Then the volumetric flow rate $Q$ for a cylindrical tube is given by the Hagen–Poiseuille equation
\begin{align}
    Q = \frac{\pi R^4}{8 \mu L} \Delta p,
\end{align}
where $R$ is the tube's radius, $\mu$ is the fluid viscosity, and $\Delta p$ is the pressure drop along the tube. We define a tube's conductivity $K = \pi R^4 / 8 \mu L$ such that $\label{eq:Pois} Q = K \Delta p.$
\begin{figure}
    \centering
    \includegraphics[scale=0.15]{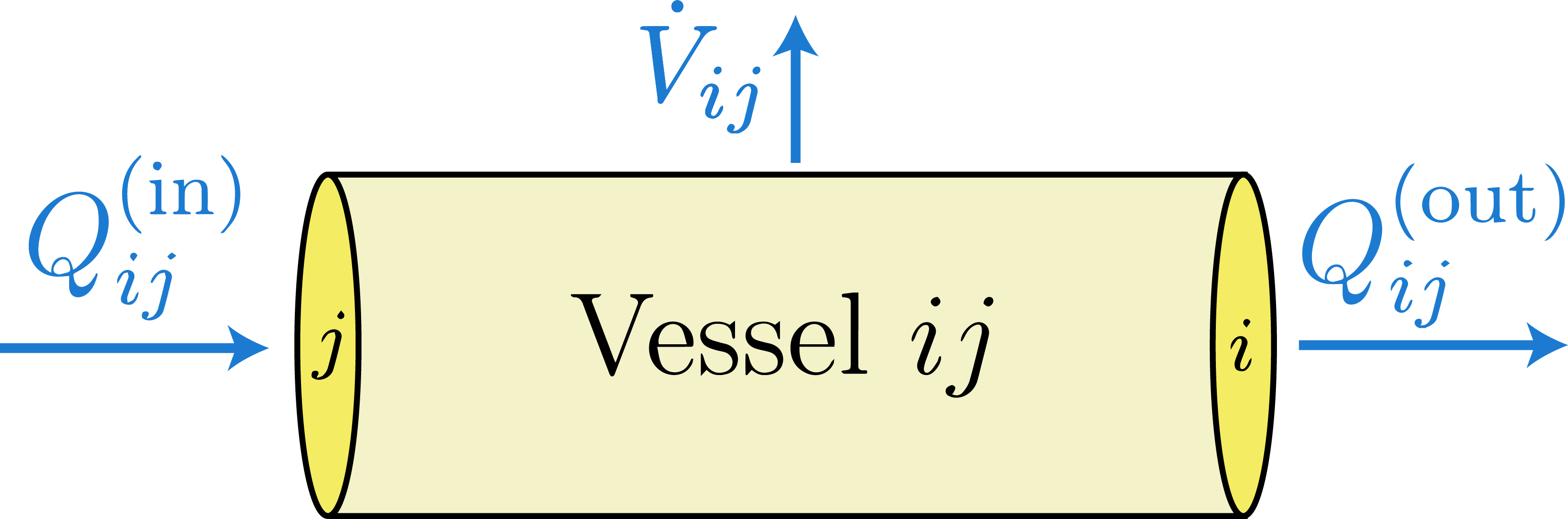}
    \caption{Fluid flows into node $j$, denoted $\Qin$, and fluid flows out of node $i$, denoted $\Qout$. This vessel is expanding, corresponding to some change in volume $\dot{V}_{ij}$. This change in volume must balance with the in-flow and out-flow to conserve volume, as given by Equation \protect\eqref{eq:flow}.}
\end{figure}
Pumping results in an inflow $Q^{(\text{in})}$ and outflow $Q^{(\text{out})}$ related by
\begin{align}
    \label{eq:flow}
    Q^{(\text{out})} = Q^{(\text{in})} - \dot{V},
\end{align}
where $V$ is the vessel volume.
By conservation of the fluid volume,
\begin{align}
    Q(x+dx) = Q(x) - \dot{A} dx,
\end{align}
where $\dot{A}$ is the cross sectional area's rate of change. Integrating we find
\begin{align}
    \label{eq:Qx}
   Q(x) &= - \dot{V} \frac{x}{L} + Q(0).
\end{align}
In a fixed tube the power dissipated by viscous forces is given by
\begin{align}
    \label{eq:p-disp}
    P = \frac{Q^2}{K}.
\end{align}
To find the power dissipated for our contracting tube, we split the tube into segments $dx$. The power dissipated in each segment is given by 
\begin{align}
    dP(x) = \frac{Q(x)^2}{KL} \, dx.
\end{align}
Using our expression for $Q(x)$ from Equation \eqref{eq:Qx}, we integrate along these segments to find the total power dissipation,
\begin{align}
   \label{eq:power}
    P &= \int^L_0 dP(x) = \frac{\lr{Q(0) - \frac{1}{2}\dot{V}}^2}{K} + \frac{1}{12K} \dot{V}^2.
\end{align}

\begin{figure}
    \centering
    \includegraphics[scale=0.15]{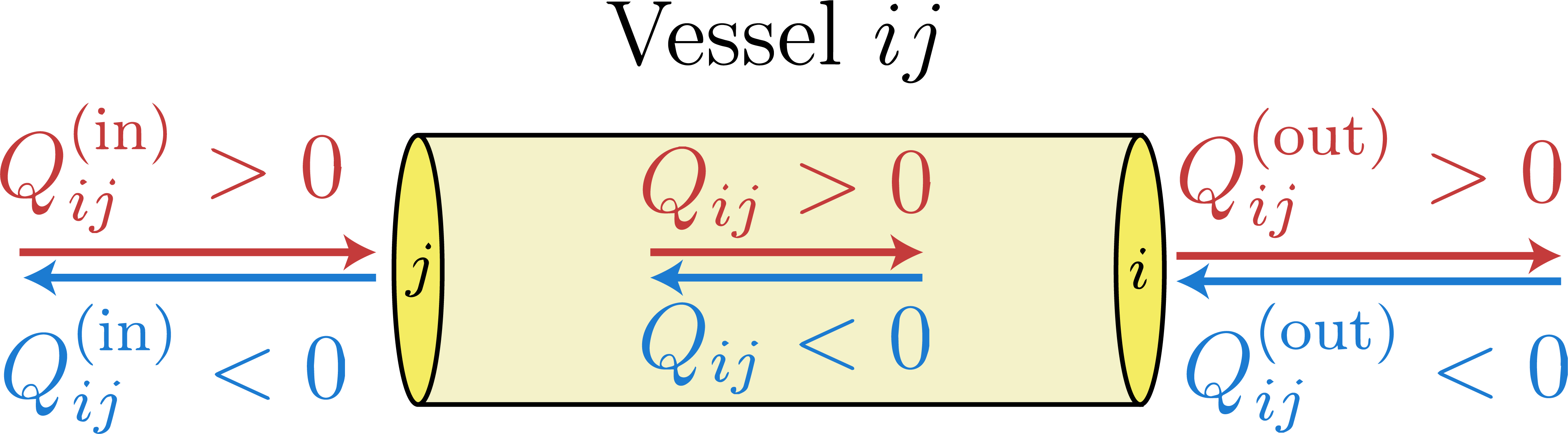}
    \caption[Flow Convention Cartoon]{A cartoon to depict our convention for flow labels. The blue font and arrows depict flow going from node $i$ to node $j$, while the red font and arrows depict flow travelling the reverse direction, from node $j$ to node $i$}
\end{figure}

We represent the network of tubes by a graph of nodes and edges, where the edge which connects node $i$ to node $j$ is denoted by $ij$. Each tube has its own volumetric inflow and outflow, labeled $\Qin$ and $\Qout$. We define
\begin{align}
    \Qin &\equiv \text{flow at node } j, \\
    \Qout &\equiv \text{flow at node } i,
\end{align}
with positive flow travelling from $j \rightarrow i$. By symmetry, we find the rules $\Qout = - Q^{(\text{in})}_{ji},$ and $\Qin = - Q^{(\text{out})}_{ji}.$ With this notation we write Equation \eqref{eq:flow} for each tube as 
\begin{align}
    \label{eq:QoutQin}
    \Qout = \Qin - \dot{V}_{ij},    
\end{align}
and using Equation \eqref{eq:power} we can express the power dissipation for each tube as
\begin{align}
    P_{ij} = \frac{Q_{ij}^2}{K_{ij}} + \frac{1}{12}\frac{\dot{V}^2}{K_{ij}}, 
\end{align}
where $Q_{ij}$ is the average flow for tube $ij$, which is obtained at the tube's midpoint. Flow conservation is then given by 
\begin{align}
    \label{eq:flow-cons}
    \sum_j Q_{ij} = \frac{1}{2} \sum_j \dot{V}_{ij},
\end{align}
while volume conservation is given by
\begin{align}
    \label{eq:vol-cons}
    \sum_{(ij)} \dot{V}_{ij} = 0.
\end{align}

\textit{Physarum}'s vessel walls behave like a visco-elastic material \cite{Oster}, which will in turn dissipate energy. To incorporate this in our minimization problem we start by considering a driven over-damped spring and look to find the proper functional to minimize which reproduces these dynamics. This system's motion is given by
\begin{align}
\label{eq:OverDamp}
    b\dot{x} = -\frac{dU(x)}{dx} + f(t),
\end{align}
where $x$ is displacement, $b$ is the damping constant, $U(x)$ a conservative potential, and $f(t)$ a driving force. We propose the functional
\begin{align}
    \label{eq:power-functional}
    \mathcal{P} = b \dot{x}^2 + 2 \frac{dU(x)}{dx}\dot{x} - 2 f(t) \dot{x}.
\end{align}
Which after minimizing does indeed reproduce a driven over-damped spring's dynamics,
\begin{align}
    \frac{\partial \mathcal{P}}{\partial \dot{x}} = 0 \implies b\dot{x} = -\frac{dU(x)}{dx} + f(t).
\end{align}
We define $U(x)$ as the potential energy due to the elastic response. To find this potential energy, which depends on the radial displacement $R$, we integrate over the surface area
\begin{align}
     U(R) = 2 \pi L \int_{R^{(0)}}^R \sigma_{\text{elastic}}(R) R \, dR.
\end{align}
With this we can write Equation \eqref{eq:power-functional} in terms of $R$ while utilizing the potential $U(R)$,
\begin{align}
     \mathcal{P} = b \dot{R}^2 + 4 \pi L R E \varepsilon \dot{R} - 2 f(t) \dot{R}.
\end{align}
Rewriting this in terms of $\dot{V}$, 
\begin{align}
    \label{eq:elastic-func}
    \mathcal{P} = \frac{b}{4 \pi^2 R^2 L^2} \dot{V}^2 + 2 E \varepsilon \dot{V} - 2 \frac{f(t)}{2 \pi R L} \dot{V}.
\end{align}
Identifying $2\pi R L$ as the vessel's surface area we see that the forcing can be rewritten as an active stress,
\begin{align}
    \sigma_{\text{active}}(t) = \frac{f(t)}{2 \pi R L}.
\end{align}
We now rewrite $b \dot{V}^2 / 4 \pi^2 R^2 L^2$ in terms of the viscosity $\eta$, which has better experimental characterization for \textit{Physarum} than the damping constant $b$.
\begin{align}
    \sigma_{\text{viscous}} &= \eta \dot{\varepsilon}, \\
    &= \frac{\eta}{R^{(0)}}\dot{R}, \\
    &= \frac{\eta}{2 \pi L R R^{(0)}} \dot{V}.
\end{align}
From this we see that the desired term to reproduce the viscous response after minimizing $\dot{V}$ is $\eta \dot{V}^2 / 4 \pi L R R^{(0)}$. Equating this with the damping term $b \dot{V}^2 / 4 \pi^2 R^2 L^2$ from Equation \eqref{eq:elastic-func}, we find that in order to reproduce the viscous response $\sigma_{\text{viscous}}$ we require
\begin{align}
    b_{\text{eff}} \equiv \frac{\pi L R}{R^{(0)}} \eta.
\end{align}
Utilizing this effective damping constant $b_{\text{eff}}$, we cast the fluid problem for a network of contracting terms into the power minimization of a functional $\Tc$:
\begin{align}
    \Tc =& \underbrace{\sum_{(ij)}\lr{ \frac{Q^2_{ij}}{K_{ij}} + \frac{1}{12} \frac{\dot{V}^2_{ij}}{K_{ij}}}}_{\text{Power Dissipation}} + \underbrace{\sum_{(ij)}\lr{\frac{\eta}{4 \pi L R R^{(0)}} \dot{V}^2 + 2 E \varepsilon \dot{V} - 2 \sigma_{ij}(t) \dot{V}}}_{\text{Coupled Elastic Vessel Walls}} \\&+ \underbrace{\sum_i 2 \lambda_i \sum_j \lr{Q_{ij} - \frac{1}{2} \dot{V}_{ij}}}_{\text{Flow Conservation}} + \underbrace{2 \mu \sum_{(ij)}\dot{V}_{ij}}_{\text{Volume Conservation}}.
\end{align}
The Lagrange multipliers $\lambda_i$ enforce flow conservation, and the Lagrange multiplier $\mu$ enforces total volume conservation. Starting with the volumetric flows, minimizing yields
\begin{align}
    \frac{\partial \Tc}{\partial Q_{ij}} = 0 \implies Q_{ij} = K_{ij}(\lambda_j - \lambda_i).
\end{align}
Here we can recognize the Lagrange multiplier $\lambda$'s role as reproducing pressure driven flow, and relabel $\lambda_i = p_i$ to yield
\begin{align}
    \label{eq:pres-driv}
    Q_{ij} = K_{ij}(p_j - p_i).
\end{align}
As such the midpoint volumetric flows are akin to Poiseuille flow. We now minimize the change in volume,
\begin{align}
    \frac{\partial \Tc}{\partial \dot{V}_{ij}} = 0 \implies 0 &= \lr{\frac{1}{6K_{ij}}+2\frac{\eta}{4\pi L R R^{(0)}}}\dot{V}_{ij} + 2E\varepsilon_{ij} - 2\sigma_{ij}(t) - \lambda_i - \lambda_j + 2 \mu, \\
    \implies \lr{\frac{1}{12K} + \frac{\eta}{4\pi L R R^{(0)}}} \dot{V}_{ij} &= -E\varepsilon_{ij}+\sigma_{ij}(t) + \frac{p_i + p_j}{2} - \mu.
\end{align}
Let's introduce another effective damping constant 
\begin{align}
    \tau = \frac{1}{12K} + \frac{\eta}{4\pi L R R^{(0)}},
\end{align}
to yield the volume dynamics 
\begin{equation}
    \label{eq:dynamics}
   \tau_{ij}\dot{V}_{ij} = -E\varepsilon_{ij}+\sigma_{ij}(t) + \frac{p_i + p_j}{2} - \mu.
\end{equation}
We can fix the multipliers using the two constraints. First, we rewrite flow conservation from Equation \eqref{eq:flow-cons} in terms of $Q_{ij} = K_{ij}(p_j-p_i)$ as
\begin{align}
    \label{eq:flow-cons-2}
    \sum_j K_{ij}(p_j-p_i) = \frac{1}{2}\sum_{j} \dot{V}_{ij}.
\end{align}
With this relation we can solve for the pressures $p_i$. To eliminate the other Lagrange multiplier, $\mu$, let's rewrite volume conservation as given by Equation \eqref{eq:vol-cons} with our dynamics
\begin{align}
    0 &= \sum_{j} \dot{V}_{ij} = \sum_{ij} \tau_{ij} \lr{-E\varepsilon_{ij} + \sigma_{ij}(t) + \frac{p_i + p_j}{2} - \mu}, \\
    \label{eq:mu}
    \implies \mu &= \frac{\sum_{ij} \tau_{ij} \lr{-E\varepsilon_{ij} + \sigma_{ij}(t) + \frac{p_i + p_j}{2}}}{\sum_{ij} \tau_{ij}}.
\end{align}
With this, we have solved the network flow problem through dynamics described by Equation \eqref{eq:dynamics} with pressures given by Equation \eqref{eq:flow-cons-2}, and external pressure $\mu$ enforcing volume conservation by Equation \eqref{eq:mu}. 
\subsection{Solute Dynamics}
We now look to implement a solute of active chemicals which flow through the network, and we do so with the advection-diffusion equation
\begin{align}
    \frac{\partial c(x,t)}{\partial t} =   \underbrace{\nabla \cdot \lr{D \nabla c}}_{\text{Diffusion}} - \underbrace{\nabla \cdot \lr{c \textbf{v}}}_{\text{Advection}},
\end{align}
where $c(x,t)$ is the concentration, $\textbf{v}$ is the flow velocity field, and $D$ is an effective diffusion constant.

For our network model we discretize this continuous equation by integrating over a control vessel with volume $V$,
\begin{align}
    \frac{d}{dt} \underbrace{\int_V  \, dV}_{=\Cc} = \int_V \nabla \cdot \lr{D \nabla c} \, dv - \int_V \nabla \cdot \lr{c \textbf{v}} \, dV,
\end{align}
while defining the solute amount $\Cc$ (in mol). Using the divergence theorem we rewrite both terms,
\begin{align}
    \frac{d\Cc}{dt} = \int_{\partial V} D \nabla c \cdot \textbf{n} \, dA - \int_{\partial V} c \textbf{v} \cdot \textbf{n} \, dA,
\end{align}
where \textbf{n} is the outward facing surface normal vector. Under the lubrication approximation, both $\textbf{v}$ and $\nabla c(x,t)$ point along the cylinder, such that they are orthogonal to $\partial V$ except at the circular ends. Then,
\begin{align}
    \label{eq:disc-advec-diff}
    \frac{\partial \Cc}{\partial t} &= \lr{D^{(\text{out})} A \nabla c^{(\text{out})} - D^{(\text{in})} A \nabla c ^{(\text{in})}} -  \underbrace{(v^{(\text{out})}Ac^{(\text{out})}}_{=J^{(\text{out})}} - \underbrace{v^{(\text{in})} A c ^{(\text{in})})}_{=J^{(\text{in})}} , 
\end{align}
where we introduce the solute in and outflows $J^{(\text{in,out})}$ with the convention
\begin{align}
    \Jin &= \begin{cases}
    \Qin c_j, & \Qin >0 \\
    \Qin c_{ij}, & \Qin < 0
    \end{cases} \\
    \Jout &= \begin{cases}
    \Qout c_{ij}, & \Qout > 0 \\
    \Qout c_{i}. & \Qout < 0 
    \end{cases}
\end{align}
Here we've identified $v^{(\text{in})}A = \Qin$ and $v^{(\text{out})}A = \Qout$, and determined $c^{(\text{in})}$ and $c^{(\text{out})}$ through our convention that positive flow goes from node $j$ to node $i$.

We define effective diffusion constants
\begin{align}
    \bar{D}^{(\text{in,out})} \equiv \frac{D^{(\text{in,out})}}{(L/2)},
\end{align}
with which we rewrite the discrete gradients in Equation \eqref{eq:disc-advec-diff} as discrete Laplacians, 
\begin{align}
    \frac{d\Cc}{dt} = J^{\inn} - J^{\out} + A \lr{\bar{D}^{\out} \Delta c^{\out} - \bar{D}^{\inn} \nabla c^{\inn} }.
\end{align}
Now let's apply this to the network's vessels and nodes. For a vessel, 
\begin{equation}
    \label{eq:vessel-dynamics}
    \frac{d\Cc_{ij}}{dt} = \Jin - \Jout + A_{ij} \lr{\bar{D}^{\inn}(c_j - c_{ij}) + \bar{D}^{\out}(c_i - c_{ij})}.
\end{equation}
For nodes, we get contributions from every vessel connected to the node. In total,
\begin{equation}
    \label{eq:node-dynamics}
    \frac{d\Cc_{i}}{dt} = \sum_j \Jout + \sum_j A_{ij} \bar{D}^{\out} (c_{ij}-c_{i}).
\end{equation}
We assume nodal volumes $V_i$ are much smaller than vessel volumes $V_{ij}$, $V_i \ll V_{ij}$. Consequently, we approximate nodal solute as small, $\Cc_i = V_i c_i \approx 0$, and as such nodal dynamics remain at a steady state $d\Cc_i / dt \approx 0$. Applying this simplifying approximation to Equation \eqref{eq:node-dynamics}, we can solve for the nodal concentrations,
\begin{align}
    0 &= \sum_j \Jout + \sum_j A_{ij} \bar{D}^{\out} (c_{ij}-c_{i}), \\
    &= \sum_{j, \Qout > 0} \Qout c_{ij} + c_i \sum_{j, \Qout < 0} \Qout + \sum_j A_{ij} \bar{D}^{\out}_{ij} c_{ij} - \sum_j A_{ij} \bar{D}^{\out} c_i, \\
    \label{eq:nodal-concs}
    \implies c_i &= \frac{\sum_j A_{ij} \bar{D}^{\out}c_{ij} + \sum_{j,\Qout > 0}\Qout c_{ij}}{\sum_j A_{ij}\bar{D}^{\out} - \sum_{j,\Qout<0}\Qout}.
\end{align}
This substantially reduces the model's computational cost.

To implement solute dynamics we need appropriate diffusion coefficients. A first order approximation would be one diffusion constant $D$ for the entire network, but previous results suggest that \textit{Physarum}'s topology is shaped by Taylor dispersion \cite{Marbach}, a phenomenon where diffusion is effectively increased by the parabolic flow field corresponding to Poiseuille flow. This effective diffusion constant is described by
\begin{align}
    D_{\text{eff}} = D \lr{1+\frac{R^2u^2}{48D^2}},
\end{align}
where $R$ is the vessel's radius, $u$ the flow velocity, and $D$ the usual diffusion constant. We can represent the flow velocity using volumetric flows,
\begin{align}
    u_{ij} = \frac{Q_{ij}}{\pi R_{ij}^2},
\end{align}
with which,
\begin{align}
    \label{eq:d-eff}
     D_{\text{eff}} = D \lr{1+\frac{Q^2}{48\pi^2 R^2_{ij} D^2}}.
\end{align}
So, for the vessel dynamics described by Equation \eqref{eq:vessel-dynamics} and the nodal concentrations described by Equation \eqref{eq:nodal-concs} we will use this effective diffusion constant. 
\subsection{Network Activity}
Activity has been implemented in many \textit{Physarum} models \cite{Julien, Oster, Radsz, Smith, Schuster} which drives the tube's rhythmic contractions and expansions. Based on the model in \cite{Julien} we implement a concentration-dependent contractile force,
\begin{align}
    F_{ij} = -F_0 \frac{\Cc_{ij}}{\Cc^{(0)}_{ij}} \lr{1-\frac{\varepsilon_{ij}}{\varepsilon_s}},
\end{align}
where $F_0$ is a force scale, $\Cc^{(0)}$ the equilibrium solute amount, $\varepsilon = (R-R^{(0)})/R^{(0)}$ is the vessel's strain, and $\varepsilon_s$ a strain scale for activation. Forcing is applied at the vessel's surface such that
\begin{align}
    F_0 = 2 \pi R L \alpha,
\end{align}
where $\alpha$ sets the stress scale. This yields the active stress
\begin{align}
    \sigma_{ij} = - \alpha \frac{\Cc_{ij}}{\Cc^{(0)}_{ij}} \lr{1-\frac{\varepsilon_{ij}}{\varepsilon_s}}.
\end{align}
Proportionality with $\Cc$ implements the signalling as the active solute coordinates contractions, while the strain contribution comes from observations of low-myosin systems typical to nonmuscle cells in which actin fibers overlap in a contracted cortex to amplify contractions \cite{hanson,janson}. 

Organisms with calcium coordinated oscillations have been shown to have stretch-activated calcium channels which produce more active ions in solute \cite{Lee} to replenish depleted chemicals. This is captured by \cite{Julien}
\begin{align}
    \frac{d\Cc_{ij}}{dt} \Bigg|_{\text{release}} = p\lr{1+\frac{\varepsilon_{ij}}{\varepsilon_{c}}} - d\Cc_{ij},
\end{align}
where $p$ is the production rate, $\varepsilon_c$ the strain scale for production, and $d$ the decay rate. Notably, we are not modelling the dynamics of all calcium in \textit{Physarum}'s vessels, but the dynamics of the active calcium. We assume that calcium is not limited within the cytoplasm, such that production and decay are not limited by availability. 
\subsection{Numerical Implementation}
Let's write down our full dynamics for the vessel's volume and solute amount. 
\begin{align}
    \label{eq:full-dynamics}
   \tau_{ij}\dot{V}_{ij} &= -E\varepsilon-\kappa\varepsilon^3 - \alpha \frac{\Cc_{ij}}{C^{(0)}_{ij}}\lr{1-\frac{\varepsilon_{ij}}{\varepsilon_s}} + \frac{p_i + p_j}{2} - \mu,\\
   \dot{\Cc}_{ij} &= \Jin - \Jout + A_{ij} \lr{\bar{D}^{\inn}_{\text{eff}}(c_j - c_{ij}) + \bar{D}^{\out}_{\text{eff}}(c_i - c_{ij})}+p\lr{1+\frac{\varepsilon_{ij}}{\varepsilon_{c}}} - d\Cc_{ij}, \\
   \text{with} \quad \tau_{ij} &= \frac{1}{12K_{ij}} + \frac{\eta}{4 \pi L R R^{(0)}} \quad \text{and} \quad \bar{D}^{(\text{in,out})}_{\text{eff}} = \bar{D}^{(\text{in,out})} \lr{1+\frac{Q_{ij}^2}{48 \pi^2 R^2_{ij} D^2}}.
\end{align}
To solve the model, we look to eliminate the unknown pressures. To start we write
\begin{align}
    \sum_j K_{ij}p_j - \sum_j K_{ij} p_i = \frac{1}{2} \sum_j \dot{V}_{ij},
\end{align}
which, considering conservation at every junction $i$, can be written as
\begin{align}
    \begin{pmatrix} \sum_j K_{1j}p_j - \sum_j K_{1j} p_1 \\ 
    \sum_j K_{2j}p_j - \sum_j K_{2j} p_2 \\
    \vdots \\
    \sum_j K_{nj}p_j - \sum_j K_{nj} p_n
    \end{pmatrix}&= \begin{pmatrix}
    -\sum_{j\neq1} K_{1,j} & K_{1,2} & \dots \\
    K_{1,2} & -\sum_{j\neq2} K_{2,j} & \dots \\ 
    \vdots & \ddots & \vdots \\
    K_{1,n} & \dots & -\sum_{j\neq n} K_{n,j}
    \end{pmatrix}
    \begin{pmatrix}
    p_1 \\
    p_2 \\
    \vdots \\
    p_n
    \end{pmatrix}, \\
    \label{eq:pres-mat}
    &= L \textbf{p}.
\end{align}
Here we have defined $L$, known as the weighted graph Laplacian by 
\begin{align}
    L_{ij} \equiv K_{ij} - \delta_{ij} \sum_n K_{in}.
\end{align}
We also want to express the right hand side of our equation for flow conservation without a sum, particularly using a vector of flows $\dot{V}$. To do this let's label the edges and vertices of a graph $G$, which is our network, with $\{e_i\}$ for the edges and $\{v_i\}$ for the vertices. We define the vector $\dot{\textbf{V}}$ by corresponding the $i$-th entry to the flow $\dot{V}$ of edge $e_i$. Then we can represent the desired sum using the adjacency matrix,
\begin{align}
    B_{ij} = \begin{cases}
    1 & \text{if } v_i \text{ is connected to } v_j, \\
    0 & \text{otherwise}.
    \end{cases}
\end{align}
Then,
\begin{align}
    \frac{1}{2}\sum_j \dot{V}_{ij} = \frac{1}{2} B \dot{\textbf{V}},
\end{align}
which together with Equation \eqref{eq:pres-mat} yields
\begin{align}
    \label{eq:flows-lin}
    L\textbf{p} = \frac{1}{2} B \dot{\textbf{V}}.
\end{align}
Although the graph Laplacian is singular due to the gauge freedom in the pressures, we can still solve for the pressures by utilizing the pseudo-inverse $L^{\dagger}$ as
\begin{align}
    \textbf{p} = \frac{1}{2} L^{\dagger} B \dot{\textbf{V}}. 
\end{align}
We now eliminate pressures, using the notation that a vector raised to some power -- like $\boldsymbol \varepsilon^3$ -- denotes element-wise exponentiation. Then the dynamics are given by
\begin{align}
     \tau \dot{\textbf{V}} &= -E \boldsymbol\varepsilon + \boldsymbol\sigma(t) + \frac{1}{2}B^{\top} \textbf{p} - \mu \textbf{1},
\end{align}
and without pressures
\begin{align}
    \label{eq:vol-lin}
    \implies \underbrace{\lr{\tau - \frac{1}{4}B^{\top}L^{\dagger}B}}_{=M}\dot{\textbf{V}} &= - E \boldsymbol\varepsilon - \kappa \boldsymbol \varepsilon^3 + \boldsymbol\sigma(t) - \mu \textbf{1},
\end{align}
where we have defined the matrix $M$. We solve for $\mu$ as
\begin{align}
    \underbrace{\textbf{1}^{\top}\dot{\textbf{V}}}_{=0} &= \textbf{1}^{\top}M^{-1} \lr{-E \boldsymbol\varepsilon - \kappa \boldsymbol \varepsilon^3 + \boldsymbol \sigma(t)} - \mu \textbf{1}^{\top}M^{-1}\textbf{1}, \\
    \label{eq:mu-lin}
    \implies \mu &= \frac{\textbf{1}^{\top}M^{-1}\lr{-E \boldsymbol \varepsilon - \kappa \boldsymbol \varepsilon^3 + \boldsymbol \sigma(t)}}{\textbf{1}^{\top} M^{-1} \textbf{1}}.
\end{align}
We are now ready to numerically solve the model as outlined in Algorithm \ref{alg:model}. For our work, all numerical solutions were found using the programming language Julia \cite{Julia}, and code is available on Github at \url{https://github.com/AdamBDionne/Physarum}. Julia's \textit{DifferentialEquations.jl} package was used to numerically solve the mode with the Runge-Kutta method of order 4 (RK4) \cite{Julia}. 

\begin{algorithm}[H]
\caption{Model Implementation}\label{alg:model}
\begin{algorithmic}[1]
\For{each time step $\Delta t$}
    \State find $\mu$ using Equation \eqref{eq:mu-lin}
    \State find $\dot{\textbf{V}}$ using Equation \eqref{eq:vol-lin}
    \State find $\textbf{Q}$, $\textbf{Q}^{(\text{in})}$, $\textbf{Q}^{(\text{in})}$ using Equations \eqref{eq:flows-lin}, \eqref{eq:Qx}
    \State find $\textbf{D}_{\text{eff}}$ using Equation \eqref{eq:d-eff}
    \State eliminate nodal dynamics using Equation \eqref{eq:nodal-concs}
    \State find $\dot{\boldsymbol \Cc}$ using Equation \eqref{eq:vessel-dynamics}
    \State use $\dot{\textbf{V}}$ and $\dot{\boldsymbol \Cc}$ to update current $\textbf{V}$ and $\boldsymbol \Cc$
\EndFor
\end{algorithmic}
\end{algorithm}

\subsection{visco-elastic Modes}
We now look to understand the stability of the in-active fluid network model. We start with a perturbation to the equilibrium vessel radii,
\begin{align}
    \textbf{R}(t) = \bar{\textbf{R}} + \textbf{r}.
\end{align}
Lets evaluate the dynamics given by Equation \eqref{eq:vol-lin} with this ansatz while only keeping terms to first order in $\textbf{r}$. Any non-bold character, say $R$, will denote a diagonal matrix. For example, $R$ is a matrix with the entries
\begin{align}
    R_{ij} = \begin{cases} \text{The radius of edge } e_i & \text{if } i = j, \\ 
    0 & \text{if } i \neq j.
    \end{cases}
\end{align}
Further, the inverse $R^{-1}$ denotes
\begin{align}
     R^{-1}_{ij} = \begin{cases} \frac{1}{\text{The radius of edge } e_i} & \text{if } i = j, \\ 
    0 & \text{if } i \neq j.
    \end{cases}
\end{align}
Also, $\textbf{1}$ will denote the vector of all ones and $\mathbb{1}$ the matrix identity. Then we perturb our dynamics given by Equation \eqref{eq:vol-lin}, 
\begin{align}
    2 \pi RL M\dot{\textbf{R}} &= - E (R-\bar{R})\bar{R}^{-1} \textbf{1} - \kappa \lr{(R-R^{(0)})\bar{R}^{-1}}^3 \textbf{1} - \mu\textbf{1}, \\
     2 \pi (\bar{R} + r) L M\dot{\textbf{r}} &= -\bar{R}^{-1} E\textbf{r} - \bar{R}^{-3} \kappa \textbf{r}^3 - \mu\textbf{1}, \\
   2 \pi \bar{R} L M \dot{\textbf{r}} &= - \bar{R}^{-1} E \textbf{r} - \mu\textbf{1} + \mathcal{O}(\textbf{r}^2).
\end{align}
Using Equation \eqref{eq:mu-lin},
\begin{align}
    \mu &= \frac{\textbf{1}^{\top} M^{-1} \lr{- E (R-\bar{R})\bar{R}^{-1} \textbf{1} - \kappa \lr{(R-R^{(0)})\bar{R}^{-1}}^3 \textbf{1}}}{\textbf{1}^{\top} M^{-1} \textbf{1}}, \\
    &= \frac{\textbf{1}^{\top} M^{-1} \lr{- \bar{R}^{-1} E \textbf{r} - \bar{R}^{-3} \kappa \textbf{r}^3}}{\textbf{1}^{\top} M^{-1} \textbf{1}}, \\
    &= \frac{\textbf{1}^{\top} M^{-1} \lr{-\bar{R}^{-1}E \textbf{r} + \mathcal{O}(\textbf{r}^3)}}{\textbf{1}^{\top} M^{-1} \textbf{1}}.
\end{align}
Combining these and dropping our higher order terms we find,
\begin{align}
    2 \pi \bar{R} L M \dot{\textbf{r}} &= - \bar{R}^{-1}E \textbf{r} - \frac{\textbf{1} \textbf{1}^{\top} M^{-1} (-\bar{R}^{-1}E \textbf{r})}{\textbf{1}^{\top} M^{-1} \textbf{1}}, \\
     &= - \lr{\mathbb{1} - \frac{\textbf{1} \textbf{1}^{\top} M^{-1}}{\textbb{1}^{\top}M^{-1}\textbf{1}}} \bar{R}^{-1}E \textbf{r}, \\
     \label{eq:mode}
      2 \pi \bar{R}^2 L M \dot{\textbf{r}} &= - \lr{\mathbb{1} - \frac{\textbf{1} \textbf{1}^{\top} M^{-1}}{\textbb{1}^{\top}M^{-1}\textbf{1}}} E \textbf{r}. 
\end{align}
This expression is approximately invariant to the vessel's geometry. This is important as we would not want any results to depend on our network decomposition. Recall that 
\begin{align}
    M = \tau - \frac{1}{4}B^{\top}L^{\dagger}B \quad \text{ where } \tau = \frac{\eta}{4 \pi L R \bar{R}} + \frac{1}{12K}.
\end{align}
For our parameters, we have
\begin{align}
    \frac{\eta}{4 \pi L R R^{(0)}} \gg \frac{1}{12K},
\end{align}
as they differ by about four orders of magnitude on average. Then, we see that the right hand side term is independent of topology while on the left hand side the diagonal terms of $M$ depend on $L^{-1}\bar{R}^{-2}$, which cancels with the $L\bar{R}^2$ factor in Equation \eqref{eq:mode}.

Let's solve for $\dot{\textbf{r}}$ by inverting the left hand side,  
\begin{align}
    \dot{\textbf{r}} = -M^{-1}\lr{2 \pi \bar{R}^2 L}^{-1} \lr{\mathbb{1} - \frac{\textbf{1} \textbf{1}^{\top} M^{-1}}{\textbb{1}^{\top}M^{-1}\textbf{1}}} E  \textbf{r}.
\end{align}
We've now uncovered a linear system, which will have solutions given by the eigenmodes
\begin{align}
    \textbf{r}(t) = e^{-\lambda t}\textbf{u},
\end{align}
where $\lambda$ and $\textbf{u}$ are the eigenvalues and eigenvectors respectively satisfying
\begin{align}
    \lambda M \textbf{u} = \lr{2 \pi \bar{R}^2 L}^{-1}\lr{\mathbb{1} - \frac{\textbf{1} \textbf{1}^{\top} M^{-1}}{\textbb{1}^{\top}M^{-1}\textbf{1}}} E \textbf{u}.
\end{align}
With this, we have found visco-elastic decay modes. The eigenvalues $\lambda_i$ denote the characteristic time for some excitation to damp out. 

These eigenmodes $\textbf{u}$ yield a discrete spectra which ranges from low to large scale excitations depending on their eigenvalue $\lambda$. To make this precise we define \textit{nodal domains} as maximally connected subsets of a function's domain for which the function does not change sign. Consider the second order differential equation\begin{align}
    \Delta \textbf{u} = - \lambda \textbf{u},
\end{align}
where $\Delta$ is the Laplace operator. In 1924 Courant proved along with Hilbert that the eigenfunction $\textbf{u}_n$, when ordered by increasing eigenvalue $\lambda_n$, separates the domain into no more than $n$ nodal domains~\cite{Courant1923}. This has been extended to the discrete case in \cite{Urschel}, which states that for a \textit{generalized Laplacian} M the eigenvector $u_n$, when ordered by increasing eigenvalue $\lambda_n$, will have at most $n$ nodal domains. We define a \textit{generalized Laplacian} as any matrix M associated to a graph $\mathcal{G}$ with vertices $\mathcal{V} = \{v_i\}$ and edges $\mathcal{E} = \{e_i\}$ such that 
\begin{align}
    M_{ij} < 0 &\quad \text{ for all } \{v_i, v_j\} \in \mathcal{E}, \\
    M_{ij} = 0 &\quad \text{ for all } \{v_i, v_j\} \not \in \mathcal{E}, i \neq j. 
\end{align}
We also need to define nodal domains for the discrete case. For our purposes, we can think of a nodal domain as a subgraph of $\mathcal{G}$ such that every value $u$ obtains on the subgraph is the same sign or $0$. 

\begin{figure}
    \centering
    \includegraphics[scale=0.7]{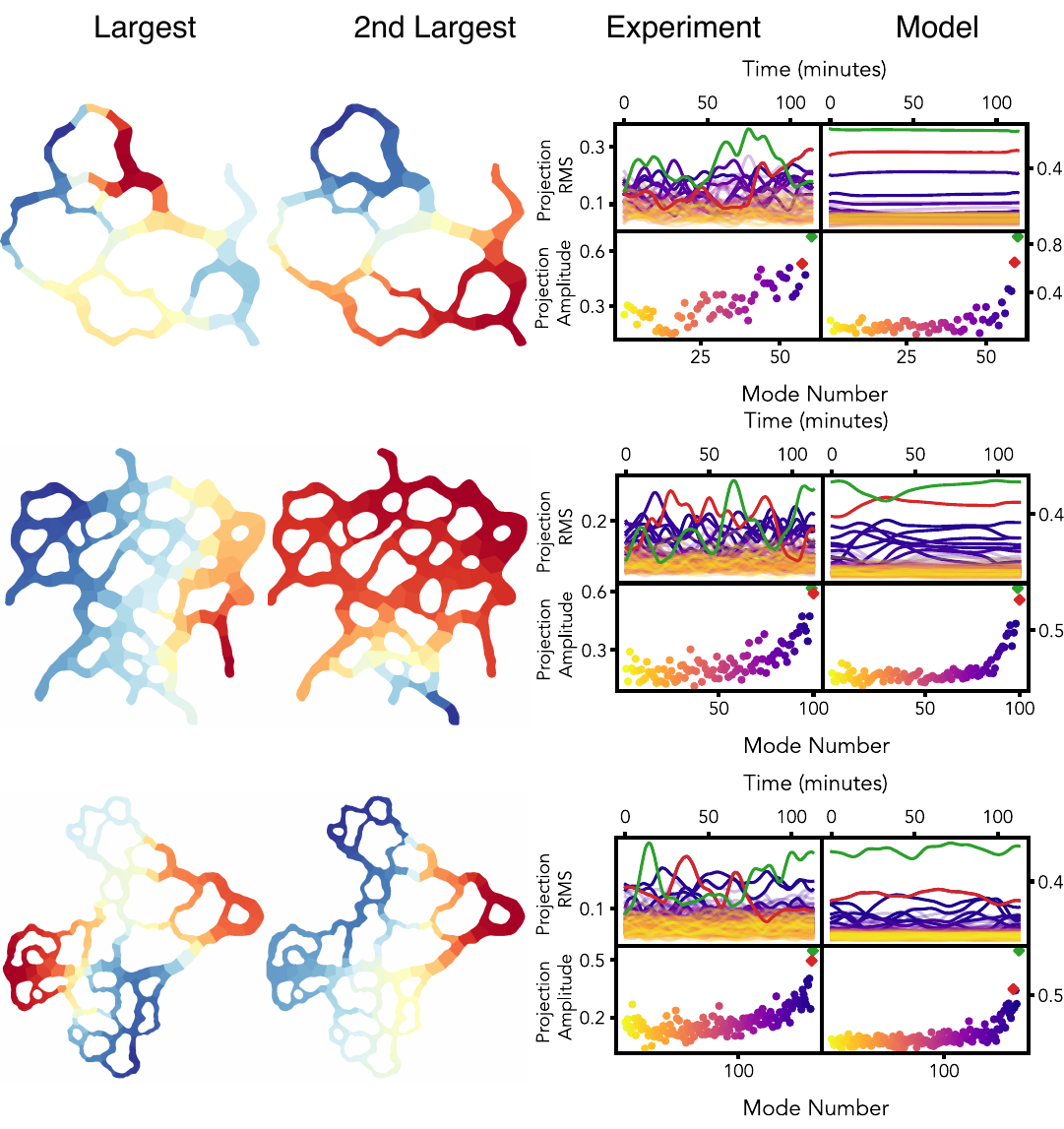}
    \caption[]{Samples one, two, and three.}
    \label{fig:modes}
\end{figure}

\begin{figure}\ContinuedFloat
    \centering
    \includegraphics[scale=0.7]{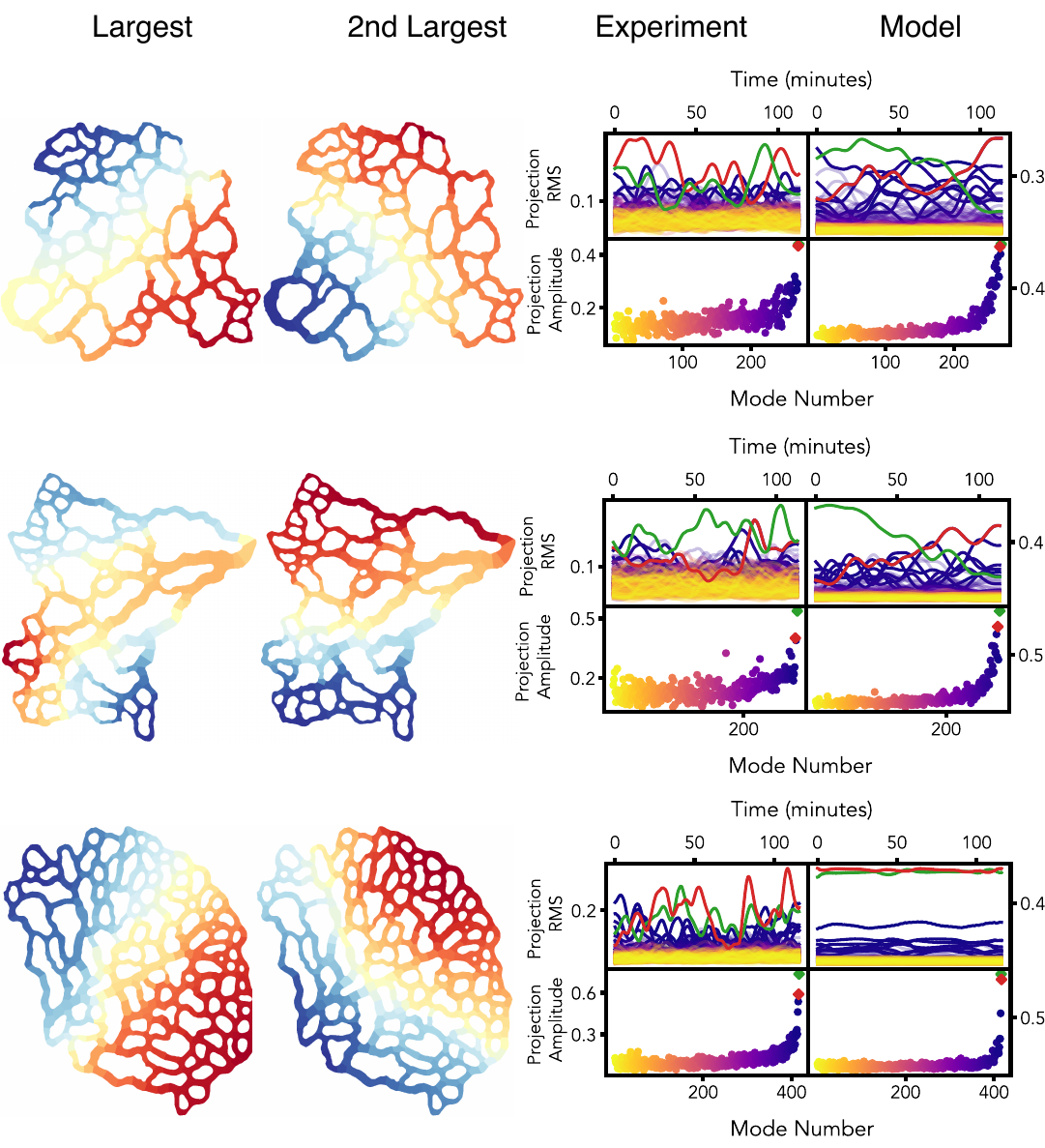}
    \caption[]{Samples four, five, and six.}
\end{figure}

\begin{figure}\ContinuedFloat
    \centering
    \includegraphics[scale=0.7]{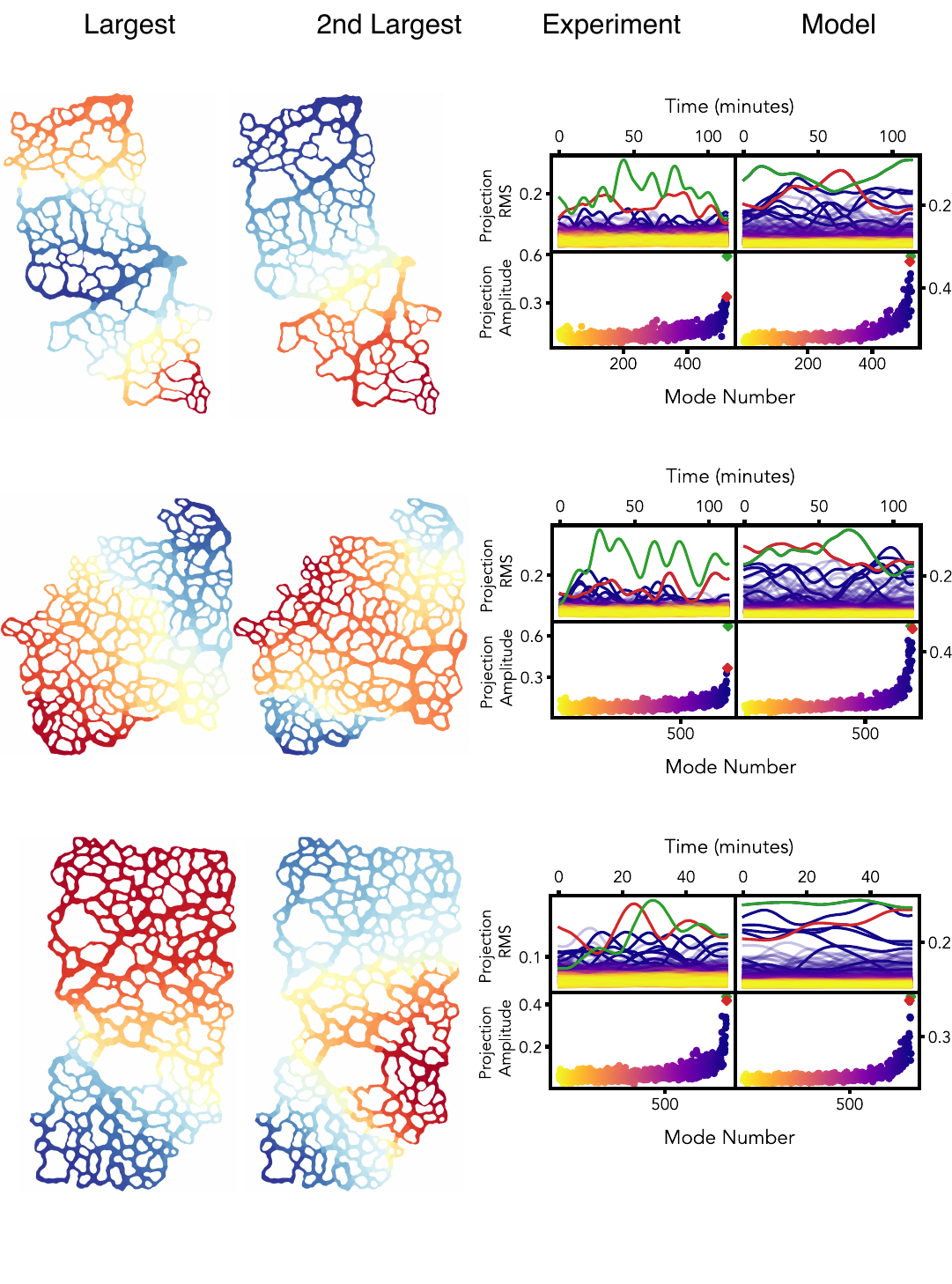}
    \caption[visco-elastic mode decomposition of \textit{Physarum}]{Samples seven, eight, and nine.}
\end{figure}

\begin{figure}
    \contcaption{The modal decomposition of our nine \textit{Physarum} samples. For each we display the two dominant modes observed on the left. The network's edges are colored to display their relative amplitude, with red dictating low and blue high. On the right we quantify the spatial coordination using the visco-elastic modes for both observed and modeled behavior. The observed behavior corresponds to the figure's third column, and the modelled behavior the fourth column. We display the projection root-mean-square (RMS) for each mode. The RMS is taken over 200 seconds. The different projections are colored based on their mode number, with yellow as low mode number and purple as high mode number. The two modes with highest excitation, $N-1$ and $N-2$ are colored in red and green respectively. Below the projection RMS we plot the projection amplitude for each mode number. We define this amplitude as the maximum achieved projection over the two hours. Throughout each sample, we find that both observed and modeled behavior exhibits organization as described by the visco-elastic modes: the behavior is encapsulated by a few dominant modes with high projection amplitude.} 
\end{figure}

To use these visco-elastic decay modes to study \textit{Physarum}'s behavior, we project the modes onto both the observed and modeled radius over time signal. Let's denote these $\textbf{r}_O(t)$ and $\textbf{r}_M(t)$ respectively. The visco-elastic decay modes are a set of eigenvectors $\textbf{u}_n$, where $n$ ranges from $1$ to the number of edges in the network. For each $n$ we find the projection
\begin{align}
    \textbf{P}_{O}(t)_n &= \textbf{r}_{O}(t) \cdot \textbf{u}_n, \\
    \textbf{P}_{M}(t)_n &= \textbf{r}_{M}(t) \cdot \textbf{u}_n.
\end{align}
These yield the projection over time for mode $n$, and for each sample's plot in Figure \ref{fig:modes} a moving root mean squared (RMS) of $\textbf{P}_{O}(t)_n$ is on the top left and a RMS of $\textbf{P}_{M}(t)_n$ is on the top right for every $n$. The curves are colored based on $n$, with low $n$ being yellow and high $n$ purple. We can see this range of colors on the bottom part of each plot, which depicts the maximum projection amplitude over time for each mode number. More precisely, on the plot's bottom left is $\max_{t}\{\textbf{P}_{O}(t)_n\}$ for all $n$ and on the plot's bottom right is $\max_{t}\{\textbf{P}_{M}(t)_n\}$ for all $n$.

\subsection{Parameter Study}

In total, there are eleven meaningful parameters we must consider. Eight of the eleven have reasonable experimentally observed values, as outlined in Table \ref{tbl:params}.
\begin{table}[t]
    \centering
    \begin{tabularx}{\textwidth}[t]{cccc}
    \hline
    \textbf{\textcolor{blue}{Parameter}} & \textbf{\textcolor{blue}{Symbol}} & \textbf{\textcolor{blue}{Value}} & \textbf{\textcolor{blue}{Source}} \\
    \hline
    Cytoplasm's viscosity & $\mu$ & $1.5 \times 10^{-3}$ Pa$\cdot$s & \cite{puchkov} \\
    Effective Stiffness & $E$ & 10 Pa & \cite{salbreux}, \cite{tazawa} \\
    Active Stress & $\alpha$ & 30 Pa & \cite{salbreux}, \cite{maitre} \\
    Diffusive Constant & $D$ & $3.33 \times 10^{-10}$ m$^2 \cdot$s$^{-1}$ & \cite{donahue}, \cite{radszuweit} \\ 
    Typical Vessel Radius & $R^*$ & 46 $\mu \cdot$m & Our data \\ 
    Calcium Production & $p$ & $R^*/192$ mol$\cdot$s$^{-1}$ & \cite{glogauer}, \cite{leepar} \\
    Calcium Decay & $d$ & $R^*/192$ s$^{-1}$ & \cite{glogauer}, \cite{leepar} \\
    Effective Viscosity & $\eta$ & $240 $ Pa$\cdot$s & \cite{tazawa}, \cite{feneberg}, \cite{saha}  \\
    Nonlinear Divergence Suppressor & $\kappa$ & Unknown & -- \\ 
    Contractile Stress & $\varepsilon_s$ & Unknown & -- \\
    Inflow Stress & $\varepsilon_c$ & Unknown & -- \\
    \end{tabularx}
    \caption[Set parameter values.]{The eight set parameter values from \protect\cite{Julien} we use for our model. Three parameter values, $\kappa$, $\varepsilon_s$, and $\varepsilon_c$, are experimentally unknown and thus will be set by us through other methods.}
    \label{tbl:params}
\end{table}

This leaves the nonlinear divergence suppressor $\kappa$, contractile stress $\varepsilon_s$, and the inflow stress $\varepsilon_c$. We set $\kappa = 2000$ Pa to suppress model non-linearities. We now turn our attention towards carefully determining $\varepsilon_s$ and $\varepsilon_c$. 

We sweep through parameter space by finding model solutions in the limit cycle regime by studying the Jacobian's eigenvalues, which we find numerically using automatic differentiation in Julia with the \textit{Zygote} package. 

There are three mechanical properties that we hope to reproduce in choosing our two characteristics strain scales: flow velocity, the contracting vessel's time period, and the vessel's maximum radial strain. We observed nine \textit{Physarum} samples and find that their mean time period is about $150$ seconds and the maximum radial strain about $0.08$. Experiments find that \textit{Physarum}'s flow velocity is on the order of $10 - 100$ $\mu$m$\cdot$s$^{-1}$ \cite{Alim}. So, we look to find parameters that best reproduce this behavior.

We study how $\varepsilon_s,\varepsilon_c$  impact six different behaviors: the contracting vessel's time period, the radial strain, the global order, the local order, the flow velocity, and the largest root-mean-squared (RMS) mode projection. To find the contracting vessel's time period, we take the cross correlation between the signal and itself. The peak of this cross correlation gives a lag between the signal and itself, which we interpret as the time period. We then take the median time period over all vessels. To find the radial strain, we find all maxima and minima of the signal. With these we compute the strain, and again find the median over all vessels. 

Global and local order are measurements of the model solution's spatial coordination. We define the local order parameter $\mathcal{O}_{\ell}$ as
\begin{align}
    \mathcal{O}_{\ell} \equiv \frac{1}{N} \left| \sum_{v_i} \sum_{e_j \text{ connected to } v_i} e^{i \phi(e_j)} \right|
\end{align}
and the global order parameter $\mathcal{O}_g$ as
\begin{align}
    \mathcal{O}_g \equiv \frac{1}{N}\left| \sum_{e_j} e^{i \phi(e_j)}\right|.
\end{align}
Here, $\{v_i\}$ denotes the vertices and $\{e_j\}$ the edges of the network. For both $N$ is a normalization factor such that $0 \leq \mathcal{O} \leq 1$. 

Flow velocities are computed from the volumetric flows, which we already compute while numerically solving the model, by dividing by the vessel's cross sectional area.
\begin{figure}
    \centering
    \includegraphics[scale=0.7]{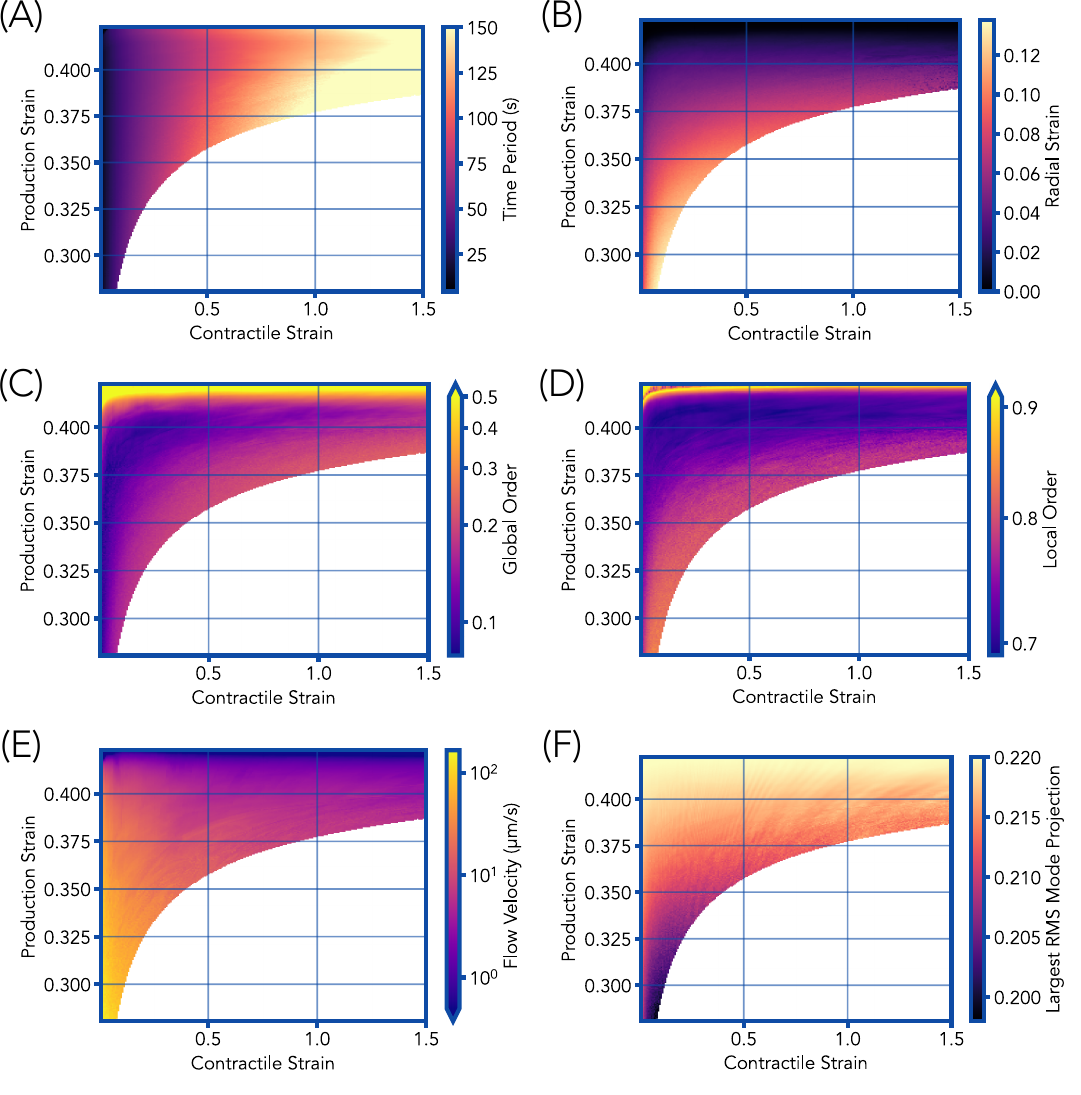}
    \caption[Phase diagrams used for parameter setting]{Six phase diagrams used to set our two fixed variables, the production and contractile strain scales. Model solution is solved for \textit{Physarum} sample one, which has 61 edges. (A) The time period of the vessel's contractions, measured in seconds. (B) The radial stress $\varepsilon$. (C) The global order $\mathcal{O}_g$ on a logarithmic scale. Bright yellow represents $\mathcal{O}_g > 0.51$, as high production strain on this plot is close to the Hopf bifurcation and is close to non-physical solutions in which the oscillations are damped to equilibrium. (D) Local order $\mathcal{O}_{\ell}$ on a logarithmic scale, bright yellow represents $\mathcal{O} > 0.91$. (E) Flow velocity, measured in $\mu$m$\cdot$s$^{-1}$, on a logarithmic scale with dark purple representing $v < 0.1$ $\mu$m$\cdot$s$^{-1}$. (F) The largest RMS of any mode projection, where the modes correspond to the scale of excitation.}
    \label{fig:phase-diagrams}
\end{figure}

From the phase diagrams in Figure \ref{fig:phase-diagrams}, we make some important qualitative observations. Firstly, the time period increases as the contractile strain $\varepsilon_c$ increases while being mostly independent of the production strain $\varepsilon_s$. Inversely, the radial strain decreases while the production strain $\varepsilon_s$ increases while being mostly independent of the contractile strain $\varepsilon_c$. Flow velocity depends mostly on the production strain, decreasing as it increases, and for $\varepsilon_s$ between $0.4$ and $0.3$ is comparable to \textit{Physarum}'s typical flow velocities. 

From this parameter search, we find that a production strain of $\varepsilon_s = 0.8$ and a contractile strain of $\varepsilon_c = 0.38$ reproduce \textit{Physarum}'s behavior quite well for this network, with a time period around $150$ seconds, a radial strain $0.07$, and flow velocities around $15$ $\mu$m$\cdot$s$^{-1}$. 

\section{Experimental methods}
We develop a methodology for observing \textit{Physarum} using bright-field microscopy. We follow methods as used in \cite{Alim, Alim2}, and \cite{Marbach}, with slight adaptations for our needs.
\begin{figure}[h]
    \centering
    \includegraphics[scale=0.2]{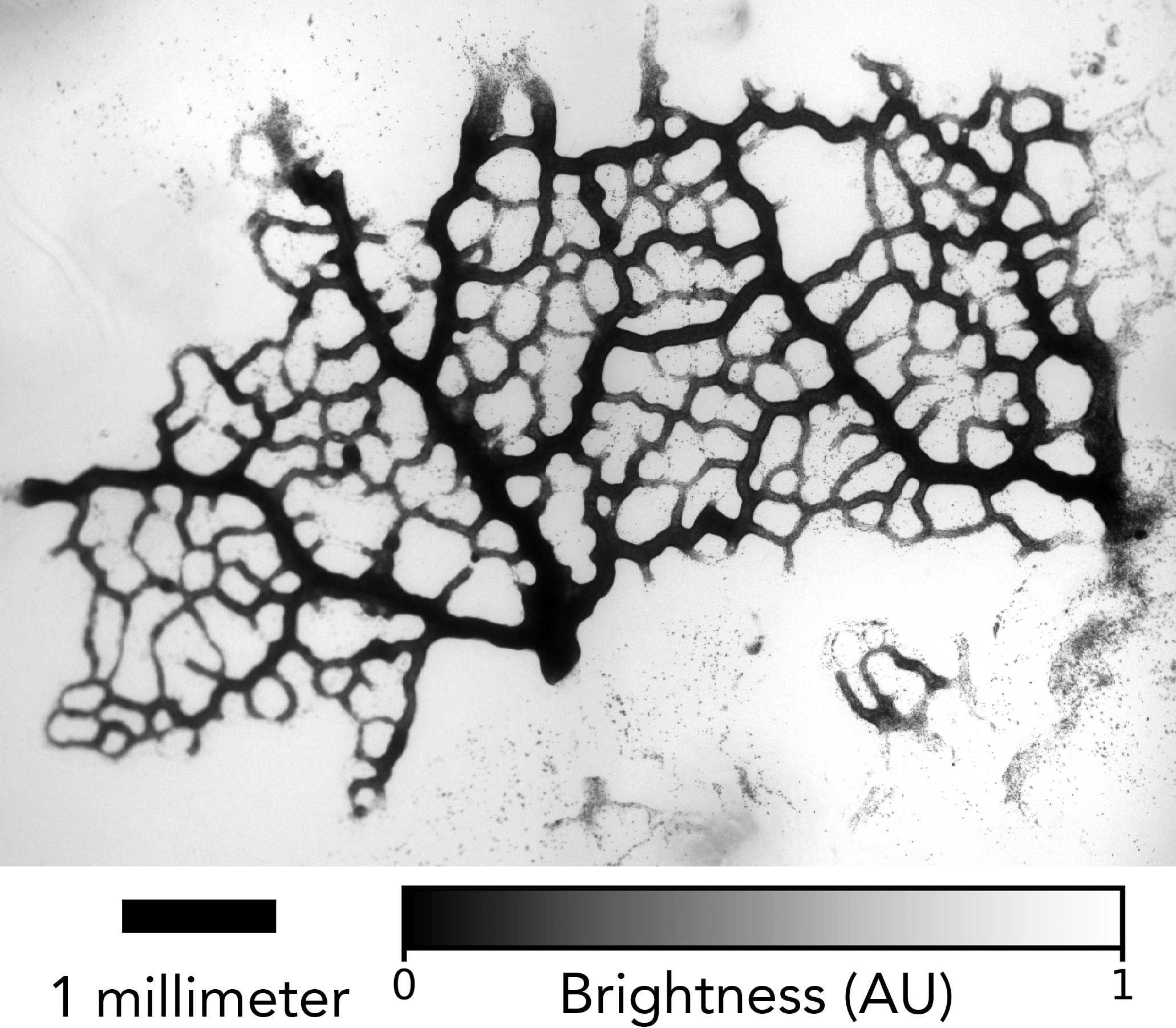}
    \caption[Raw microscopy image of \textit{Physarum}]{A raw microscopy image of \textit{Physarum}. The scale bar on the left is one millimeter, while the color bar represents brightness in arbitrary units. To measure the radius of each vessel in this image, we look to remove noise and anything not part of the network.}
    \label{fig:raw-phys}
\end{figure}

We acquired plasmodia of \textit{Physarum polycephalum} from Carolina Biological Supply. A new colony of \textit{Physarum} is cultured by cutting a segment of \textit{Physarum} using a sterilized scalpel and transferring it to a fresh media plate. We use 100 x 15 mm media plates with non-nutritive agar 1.5\% wt/vol (Carolina Biological Supply). Several oat flakes (Quaker Oats Co.) are added to the plate's center where the \textit{Physarum} segment is introduced, and several other oat flakes are scattered throughout to promote growth. Newly cultured \textit{Physarum} are stored in an unsealed but closed box such that little light enters, and samples are temperature controlled by a heating pad set to $21^{\circ}$ C (Zoo Med Laboratories, Inc.). After two to three days, we image the sample by cutting out a smaller segment placed in a 35 x 9 mm media plate. A media plate is used rather than a glass slide to minimize surface deformations of the agar, which cause the sample to move throughout imaging. A scalpel is used to scrape away growing fans and oat flakes such that imaging will capture a full network. 

\subsection{Microscopy}
We illuminate the \textit{Physarum} sample in bright-field using an inverted optical microscope (Nikon Eclipse Ti2 equipped with a Nikon DS-Qi2 monochrome camera), and image using a 4x Air (N.A. 0.13) objective lens. An image is then captured every five seconds over two hours. Given \textit{Physarum}'s typical contraction period of two to three minutes, this is adequate temporal resolution to capture the organism's behavior over many periods. 

\subsection{Post Processing}
Our goal is to extract a network representation of \textit{Physarum}, along with its spatial and temporal coordination. We start by decomposing \textit{Physarum} into a graph of nodes and edges. Using this graph we develop a methodology for measuring the radius of each \textit{Physarum} vessel for a still frame image. Doing this for every frame collected yields the temporal coordination, a radius over time signal. Finally, with some processing we extract the phase of each edge over time which yields the spatial coordination.
\begin{figure}[h]
    \centering
    \includegraphics[scale=0.07]{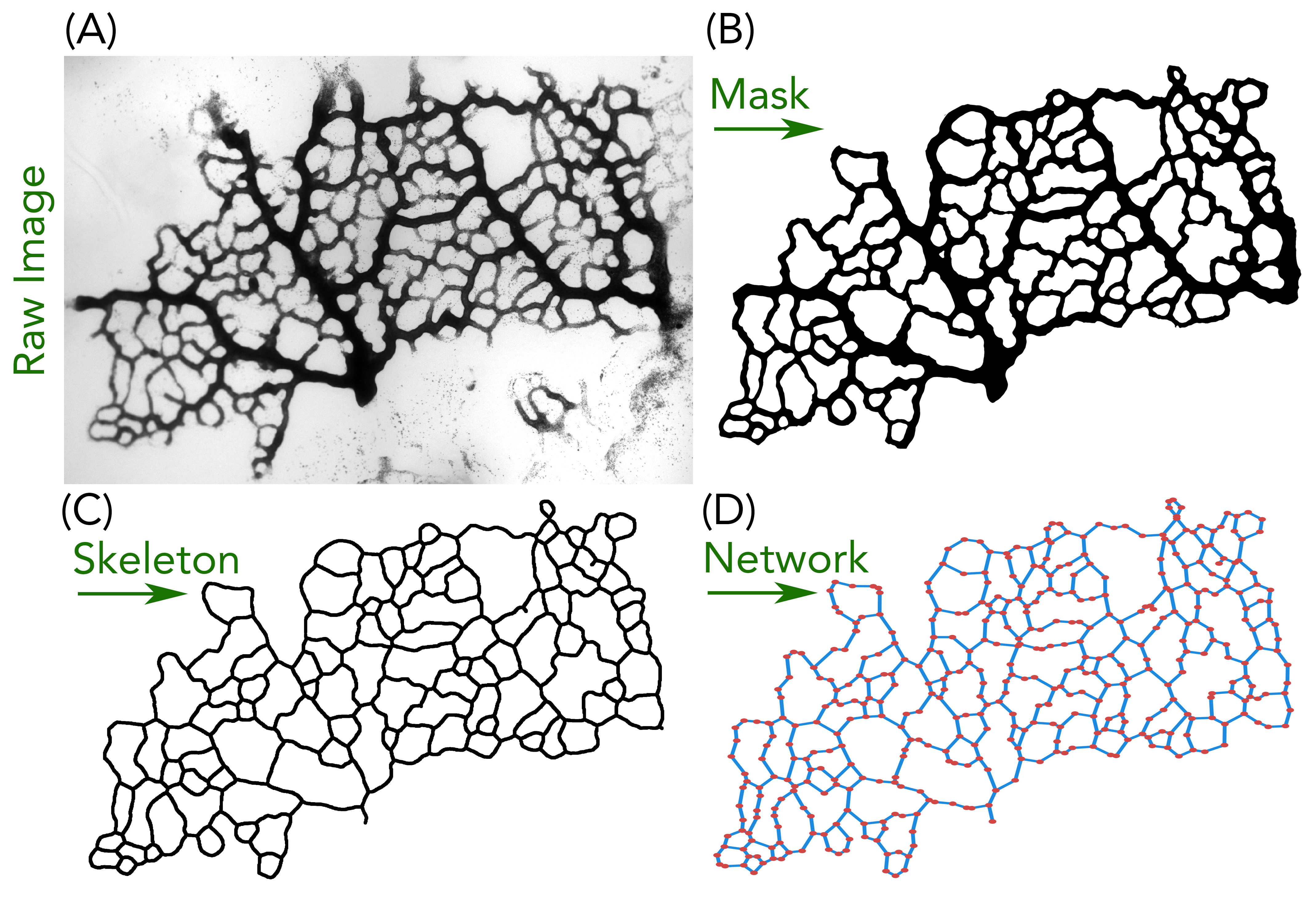}
    \caption[Decomposing \textit{Physarum} into a Network]{Our process for decomposing \textit{Physarum} into a network. (A) The raw microscopy image of \textit{Physarum} before processing. (B) A masked version of the raw image in which I've removed noise and anything not on the network. (C) A one-pixel wide curve, called a skeleton, that traces through the middle of (B). The curve's width is exaggerated for visual clarity. (D) The skeleton decomposed into a network of nodes and edges.}
    \label{fig:phys-network}
\end{figure}
We create a masking image using Adobe Photoshop to remove unrelated background, such as segments of \textit{Physarum} that are disjoint from the network. We also remove from the mask vessels that are pruned during the experiment. This maks is then converted into a one-pixel wide skeleton curve using MATLAB's ``bwskel''. Following Algorithm \ref{alg:make-network} we convert the skeleton into a network of nodes and edges. We subdivide edges to better approximate the network by Algorithm \ref{alg:subdivide-network}. 
\begin{algorithm}[H]
\caption{Network Decomposition}\label{alg:make-network}
\begin{algorithmic}[1]
\ForAll{pixels in skeleton}
    \If{adjacent to $3$ pixels in skeleton}
        \State mark pixel as node
    \EndIf
\EndFor
\ForAll{nodes}
    \State get pixel position of node
    \Repeat
    \State mark current pixel as traveled
    \State travel to adjacent untraveled pixel
    \Until{reach a node}
    \State mark connection between these two nodes as an edge
\EndFor
\end{algorithmic}
\end{algorithm}

\begin{algorithm}[H]
\caption{Subdividing Edges to Reflect \textit{Physarum} Topology}\label{alg:subdivide-network}
\begin{algorithmic}[1]
\Repeat
    \ForAll{edges in network}
        \ForAll{pixels on edge}
            \State find minimum distance from pixel to skeleton
            \State update maximum distance found so far 
        \EndFor
        \If{maximum distance found is above the threshold}
            \State subdivide edge at the pixel where the maximum was realized
        \EndIf
    \EndFor
\Until{all edges below threshold}
\end{algorithmic}
\end{algorithm}
\begin{figure}[h!]
    \centering
    \includegraphics[scale=0.07]{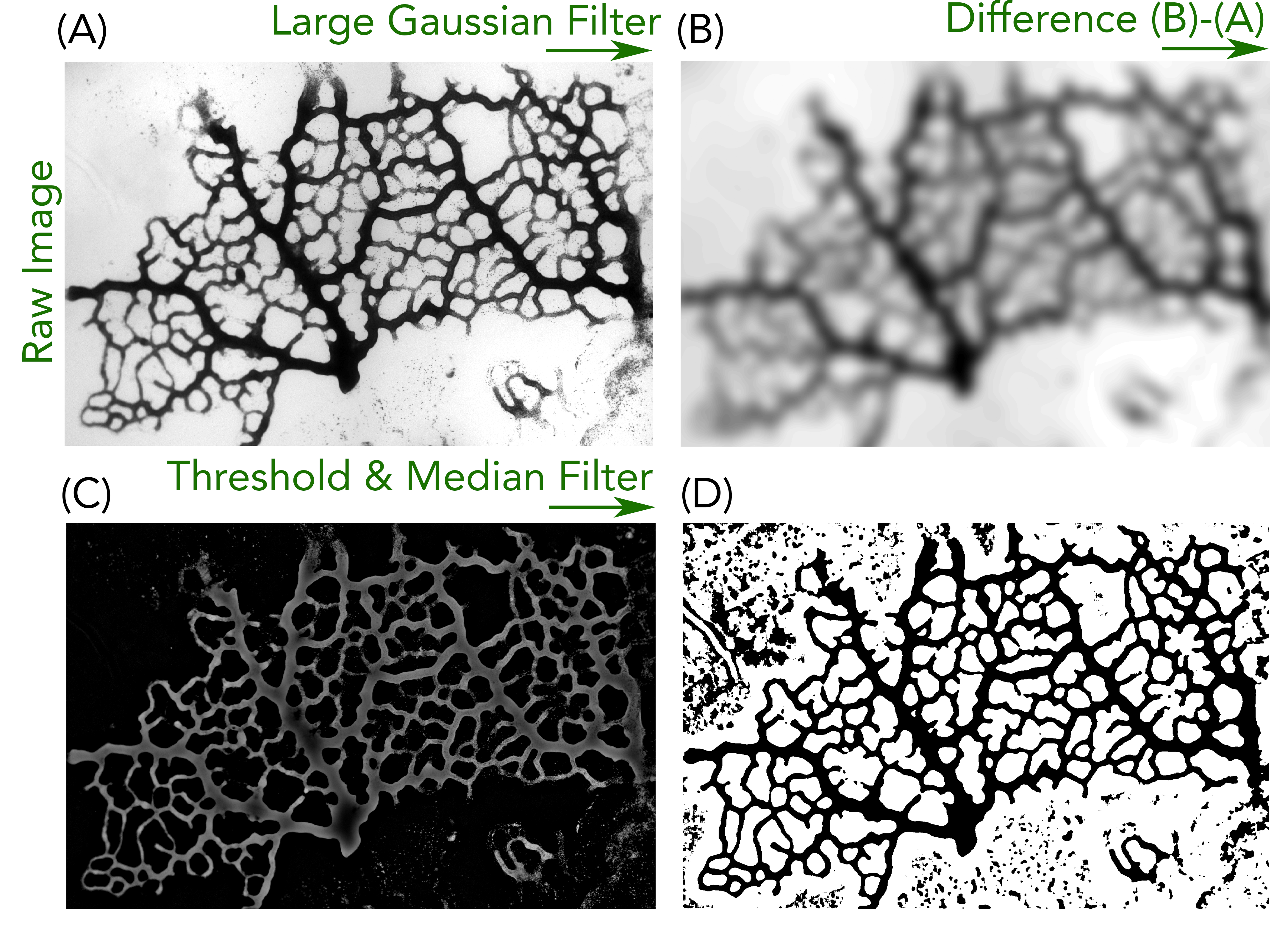}
    \caption[Processing raw microscopy image to measure vessel radii.]{Our image processing scheme used to measure vessel radii. (A) The raw microscopy image of \textit{Physarum}. (B) The raw microscopy image after applying a large Gaussian filter, $50 \times 50$ pixels. The filter size is fixed such that it is larger than a vessel but smaller than the network topology. (C) The difference between images (B) and (A), which leaves us with a denoised \textit{Physarum} network. (D) Image (C) after applying a threshold to binarize the image, and a median filter to clean up small noise from thresholding. Filter size is twenty by twenty pixels, which is smaller than vessels such as to not change vessel sizes.}
    \label{fig:raw-proc-scheme}
\end{figure}
To filter noise we apply a large Gaussian filter, $50 \times 50$ pixels, to the raw image and subtract this filtered image from the raw image. We then binarize this image, and apply a median filter. This process is depicted in Figure \ref{fig:raw-proc-scheme}.

\begin{figure}[h!]
    \centering
    \includegraphics[scale=0.28]{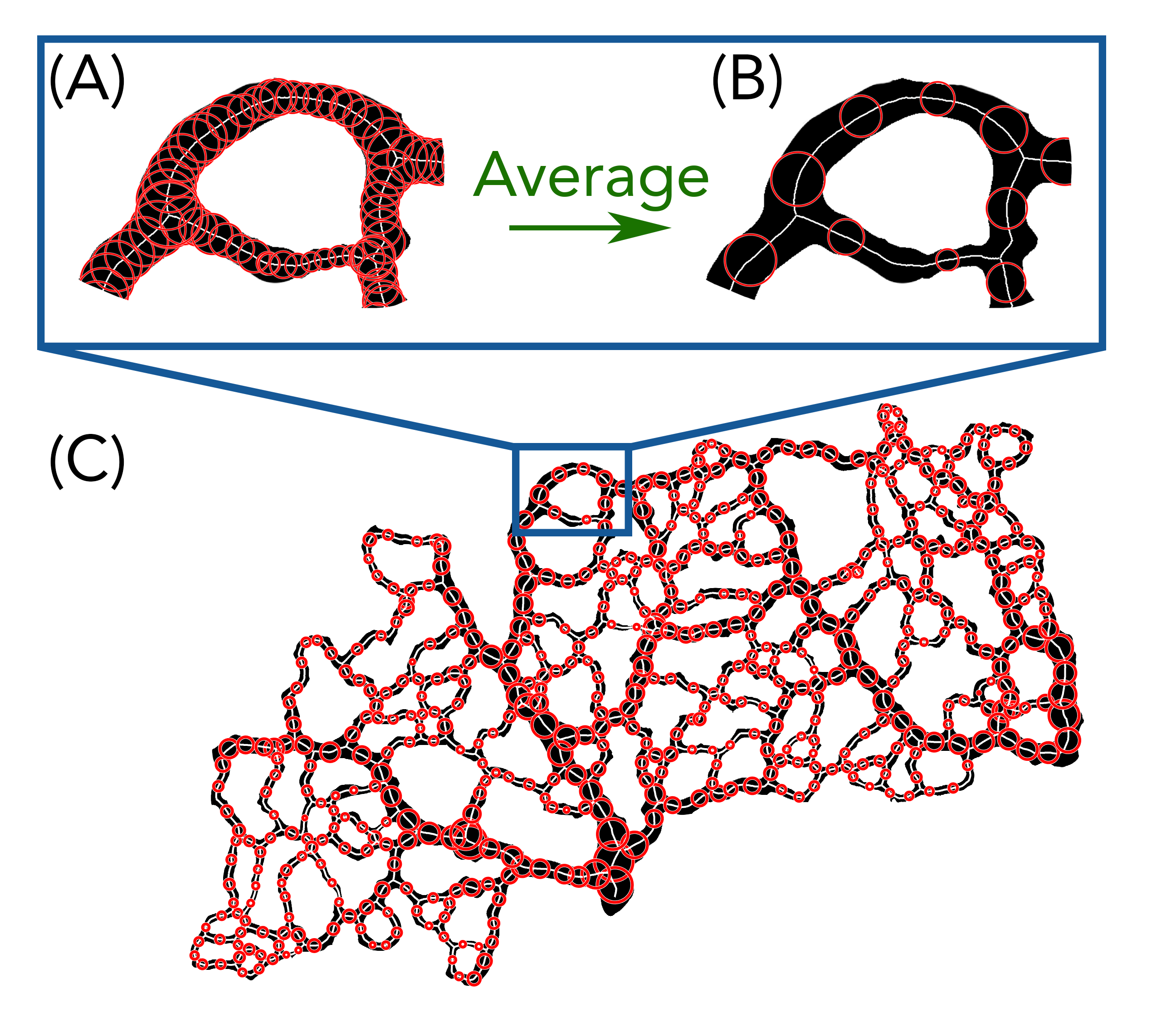}
    \caption[Example measurement of vessel radii for still frame of \textit{Physarum}.]{Example measurement of vessel radii for a still frame of \textit{Physarum}. (A) A segment of the network, as highlighted by the blue box within (C), for which we find the largest circle within the vessels for each pixel along the white curve, which is the skeleton found earlier. For visual clarity, only every tenth circle is depicted by a red circle. (B) The resulting vessel radius obtained by taking an average of all circle radii along the white curve belonging to each edge. (C) A still frame of \textit{Physarum} where each vessel's measured radius is depicted by a red circle. }
    \label{fig:meas-radii}
\end{figure}

Lastly, with this processed microscopy image we measure the radius of each vessel in the network. Here we use a methodology introduced in \cite{Bauerle}. Implemented by Algorithm \ref{alg:measure-radii} and visualized in Figure \ref{fig:meas-radii}, for each edge we consider the curve defined by the skeleton between its two nodes. For each pixel on this curve, we find the largest circle centered on the pixel within the \textit{Physarum}. The radius is then defined as the average circle radius found over all pixels on the curve. 

\begin{algorithm}[H]
\caption{Measuring a Vessel's Radius}\label{alg:measure-radii}
\begin{algorithmic}[1]
\ForAll{edges in network}
    \ForAll{pixels on skeleton curve between edge's two nodes}
        \State find the largest circle centered on this pixel that does not intersect the image's background
    \EndFor
    \State save edge's radius as average over all radii measured along skeleton curve
\EndFor
\end{algorithmic}
\end{algorithm}

\begin{figure}[h!]
    \centering
    \includegraphics[scale=0.4]{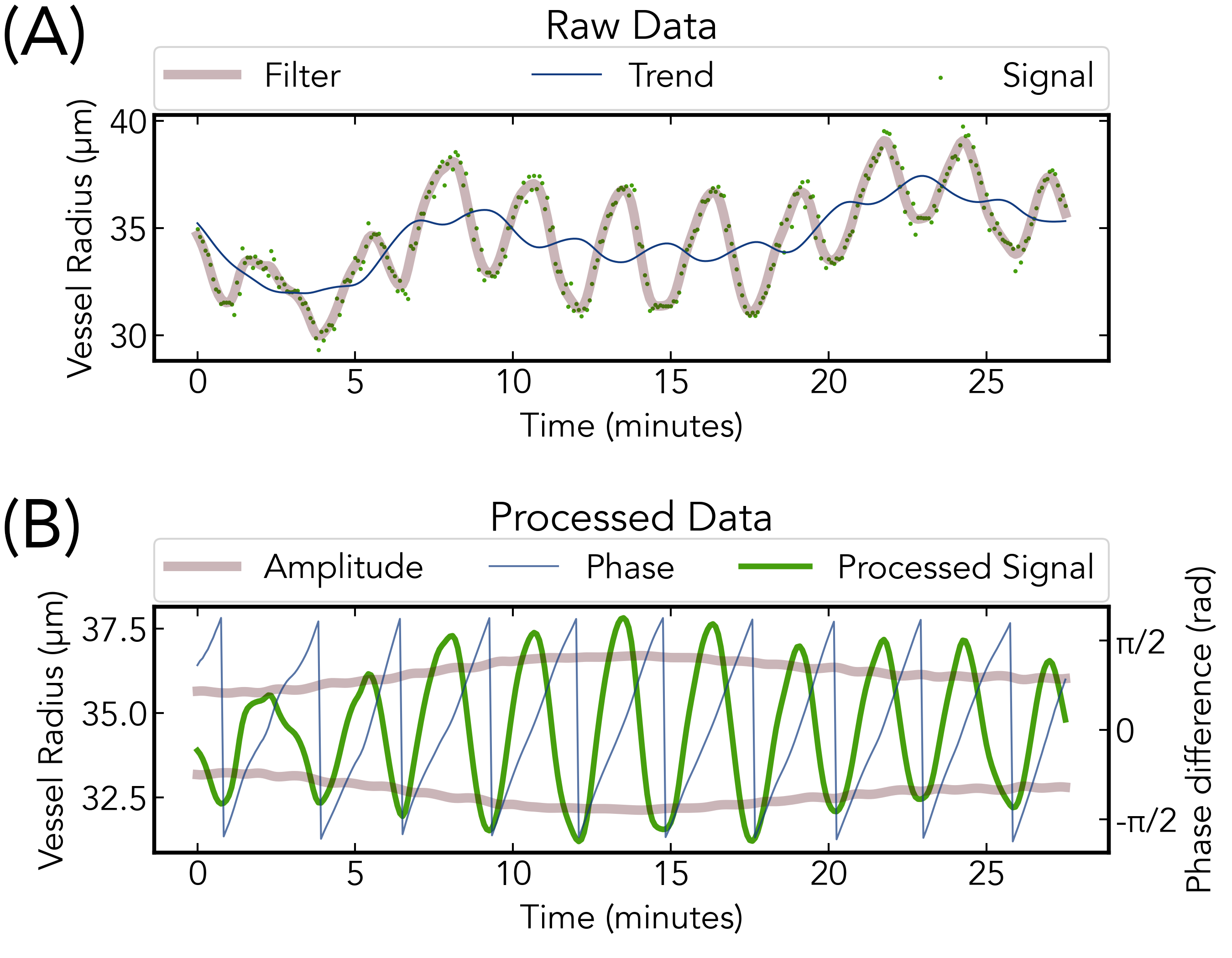}
    \caption[Processing radius over time signal for a vessel within the \textit{Physarum} network.]{Processing radius over time signal for a vessel within the \textit{Physarum} network. (A) The raw radius over time signal measured. The red curve is a Gaussian filter with window 100 seconds, which is around one half-cycle, used to smooth out the signal. The blue curve is the signal's trend, a moving average with window 200 seconds, which reflects \textit{Physarum}'s typical time period. The green points are the measured radii over time. (B) The processed radius over time signal, found by removing the blue trend above from the red filter above. The red curve is a moving RMS average over one period. The blue curve is the signal's phase, found using the Hilbert transform. The green curve is the processed signal.}
    \label{fig:radii-proces}
\end{figure}

To find the radius over time signal for each vessel we repeat this process for every microscopy image taken. An example raw signal is depicted in Figure \ref{fig:radii-proces}. We process this signal, following \cite{Bauerle}'s methodology, by applying a Gaussian filter and then de-trending the signal. The filter is over 100 seconds, around one half-cycle, and we find the trend over 200 seconds, a little over one period. The filter works to smooth biological and imaging noise, and the trend removes error from the sample or \textit{Physarum} vessel moving such that the skeleton curve is no longer centered. Taken together this produces a clear sinusoidal signal, as seen in Figure \ref{fig:radii-proces}.

\section{Nutrient dispersal simulations}
To simulate nutrient dispersal we assume that the nutrients we investigate are  are large enough that the rate of diffusion is dominated by advection. This is based in experimental observation, as under the microscope we can see particles, perhaps pieces of oat flake, being carried by the flow's current. Throughout the simulation, we describe each particle with two parameters: the current edge within the network it is in, and its current position along this edge. These parameters determine the particle's position within the network, and this position is updated by advection.

To describe advection, we need the flow velocity at any location within the network. We know from Equation \ref{eq:Qx} that the volumetric flow $Q_{ij}$ varies linearly across the vessel,
\begin{align}
    Q_{ij}(x) = -\dot{V}_{ij} \frac{x}{L_{ij}} + Q^{(\text{in})}_{ij}.
\end{align}
So we use this to determine the volumetric flow $Q_{ij}$ at the particle's position. We then find the flow velocity by dividing by the vessel's surface area,
\begin{align}
    \label{eq:flowvel}
    v_{ij}(x) = \frac{Q_{ij}(x)}{\pi R_{ij}^2}.
\end{align}
With this flow velocity we update the particle's position by numerically integrating,
\begin{align}
    x(t) = \int v(t) \, dt.
\end{align}
We continue to update the particle's position by integration until the particle reaches a node. The particle will then choose a new vessel to enter based on a stochastic process. First, we determine all vessels connected to the node that have flow travelling out of the node such that the direction of flow is preserved. This may yield multiple valid vessels, and we choose the one to travel to based on a weighted probability described by each vessel's volumetric flow magnitude $|Q_{ij}|$. The nutrient dispersal simulation is summarized in Algorithm \ref{alg:nutdisp}. 

\begin{algorithm}[H]
\caption{Nutrient Dispersal Simulation}\label{alg:nutdisp}
\begin{algorithmic}[1]
\For{each small time step $\Delta t$}
\ForAll{particles being simulated}
\State find $Q$ at particle's position using Equation \eqref{eq:Qx} 
\State update particles position using $v$ given by \eqref{eq:flowvel}
\If{particle's new position is within a node}
\State find all edges connected to node that preserve flow direction 
\State find $|Q|_i$ for each such valid edge
\State compute $\sum |Q|_i$ for valid edges
\State give each edge probability of traversal $p_i = |Q|_i /\sum |Q|_i$
\State choose edge to place particle in based on probabilities $p_i$
\EndIf
\EndFor
\EndFor
\end{algorithmic}
\end{algorithm}

For our simulation, a particle's position at time $t$ is given by two parameters, which vessel it is in and how far along the vessel. Given these two values at some time $t$ and their starting values, we compute the displacement by adding two new vertices to the network. These new vertices will be placed at the particle's starting position and the particle's position at time $t$. Then, we compute the displacement as the total weight of the shortest path between the two vertices we add, where each edge's weight is its length $L$. The shortest path is computed by Dijkstra's algorithm. 

To simulate nutrient dispersal using experimental flows, we estimate the volumetric flows $Q_{\text{obs}}$ using our radius over time signal. We convert this signal into a volume over time signal using the measured vessel length, and then estimate $\dot{V}$ at time step $t$ by taking the finite difference,
\begin{align}
    \dot{V}(t) \approx \frac{V(t+1) - V(t)}{\Delta t}. 
\end{align}
Once we have estimated $\dot{V}$ over time, we find $Q_{\text{obs}}$ by solving for the pressures and then finding the pressure driven flows.

\subsection{Transport Simulation Results}

We simulate nutrient dispersal for all nine \textit{Physarum} samples, as shown in Figure \ref{fig:dispersal}. For each, we simulate dispersal using the flows generated by the model and effectively observed flows. Plotting these simulations on a log-log plot, we see that nutrient dispersal for both observed and modeled flows follow a power law. The average power law for simulations based on modeled flows is $\gamma = 0.33$, and the average power law for simulations based on observed flows is $\gamma = 0.41$. Both $\gamma$ are less than $1/2$, and as such we find that dispersal exhibits sub-diffusive behavior. 

Note that finding $\gamma < 1/2$ does \textit{not} mean that the nutrients are being dispersed more slowly than they would by diffusion. For the particles we are simulating, their diffusive constants are small such that advection dominates. 
\newpage

\begin{figure}[h!]
    \centering
\includegraphics[width=\textwidth]{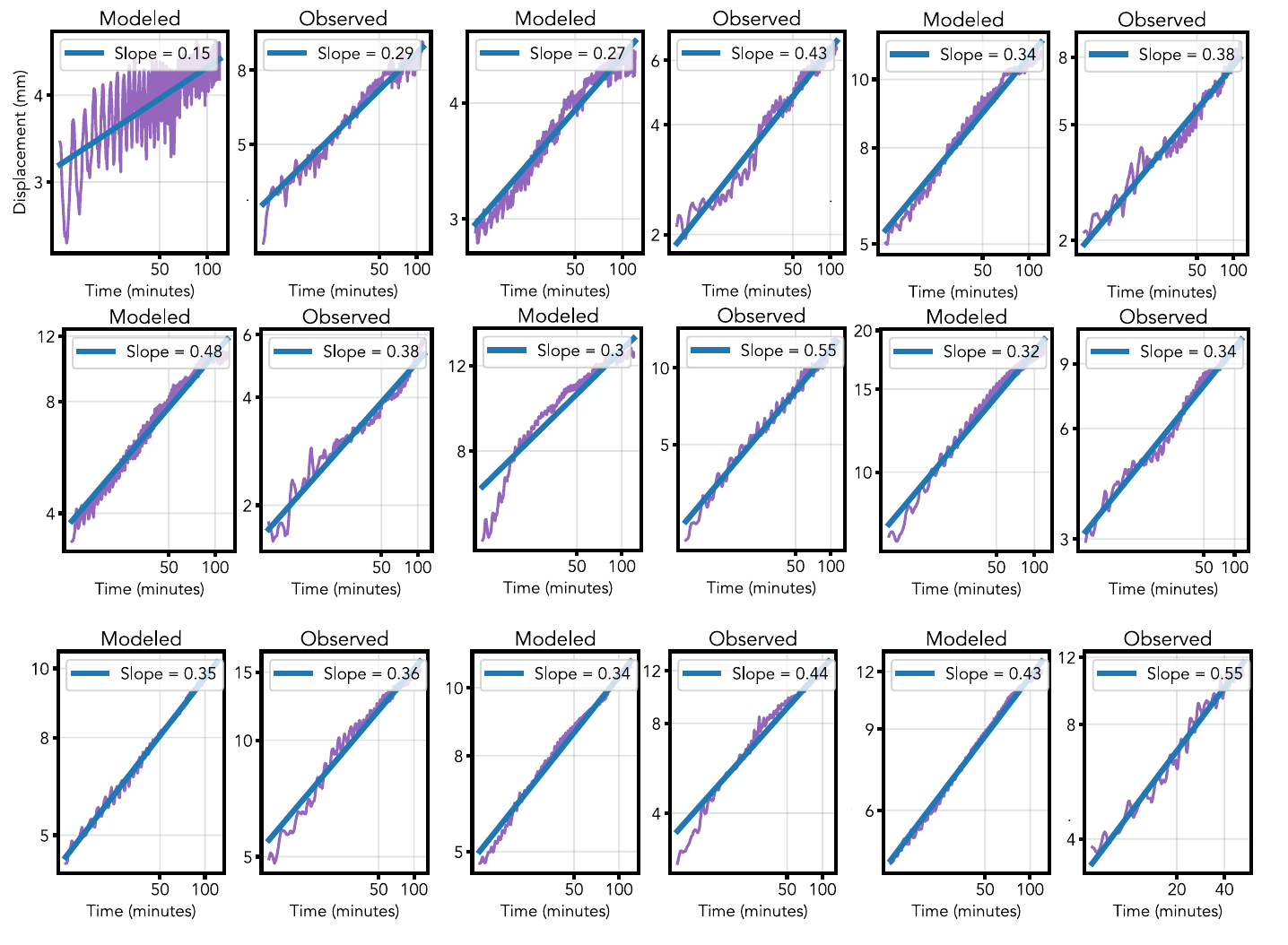}
    \caption[Nutrient Dispersal Simulations on \textit{Physarum} Networks]{Nutrient dispersal simulations on our nine \textit{Physarum} networks, comparing modeled and observed flow velocities. For each we plot the average displacement over all particles simulated, which is three particles per edge in the network, through two hours. The plots are all on log-log axes, with a line of best fit. The line's slope is denoted by the legend, which corresponds to a power-law which characterizes the dispersal behavior.}
    \label{fig:dispersal}
\end{figure}

We also simulate dispersal on a network that excites only two dominant modes. We look to investigate which modes and what phase shift $\phi$ between them will optimize dispersal. Both modes are driven with a sinusoid at the observed time period for \textit{Physarum}, with one mode given the phase shift $\phi$. The mode's amplitude is set to match the maximum radial strain observed in \textit{Physarum}. This yields the radius over time signal, which is used to find the flows using the same methodology for finding observed flows. 

Given a mode pair, we find the optimal phase difference $\phi$ for nutrient dispersal by trying all phases $[-\pi, \pi]$. Doing this on a \textit{Physarum} network for every possible pair between the 20 modes with the largest excitation size, we find that a phase shift $\phi = \pi/2$ optimizes nutrient dispersal (as long as we are not overlapping a mode with itself, then $\phi = 0$ optimizes dispersal). This is seen in Figure \ref{fig:modedisp}. Further, we find that the modes with the largest excitation will optimize dispersal. The pair which disperses the most nutrient is overlapping the two most dominant modes, $N=1$ and $N=2$, with a $\pi/2$ phase shift. 

\begin{figure}
    \centering
    \includegraphics[scale=0.85]{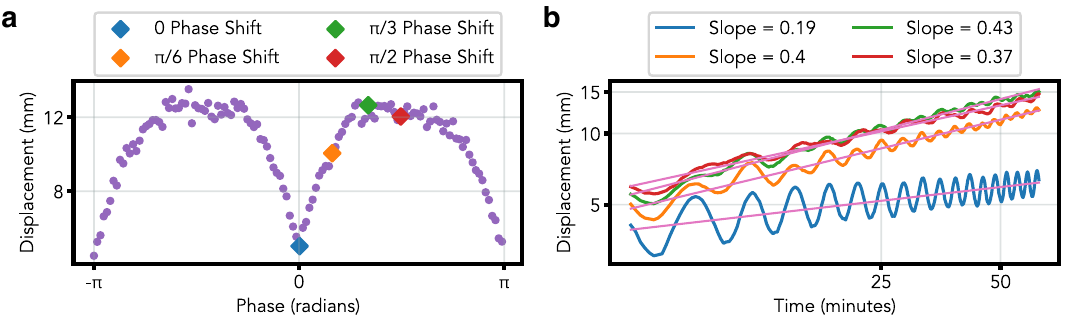}
     \caption[]{(Continued on next page)}
    \label{fig:modedisp}
\end{figure}

\begin{figure}\ContinuedFloat
    \centering
    \caption[visco-elastic decay modes at a $pi/2$ phase shift drive nutrient dispersal]{Investigating visco-elastic decay modes impact on nutrient dispersal. (A) An example demonstrating how we determine the optimal phase shift $\phi$ for two overlapped mode. Here we have overlapped the two modes with the largest excitation, $N-1$ and $N-2$, for phases within $[-\pi,\pi]$. For each phase $\phi$ we plot its maximum displacement at the end of the simulation. We find that there is a flat peak around $\pm \pi/2$, the optimal shift for transport. Four different phases are highlighted with diamonds. (B) Displacement over time plots for the four highlighted phases. We see that both $\pi/2$ and $\pi/3$ give similar results.}
    \label{fig:my_label}
\end{figure}

\bibliography{suppbib.bib}